\documentclass[twocolumn,nofootinbib,superscriptaddress,floatfix,noeprint,longbibliography]{revtex4-2}

\usepackage{graphicx,tabularx,makecell}
\usepackage{amsmath,amsfonts,amssymb}
\usepackage{mathtools}
\usepackage{bm,dsfont}
\usepackage{aligned-overset}
\usepackage[colorlinks=true, citecolor=blue, linkcolor=red, urlcolor=blue]{hyperref}
\usepackage[all]{hypcap}
\usepackage{hhline}

\graphicspath{{figures/}}

\DeclareMathOperator{\sgn}{sgn}

\begin{document}

\title{Topological Andreev Rectification}

\author{Pok Man Tam}
\thanks{These two authors contributed equally.}
\affiliation{Department of Physics and Astronomy, University of Pennsylvania, Philadelphia PA 19104}
\author{Christophe De Beule}
\thanks{These two authors contributed equally.}
\affiliation{Department of Physics and Materials Science, University of Luxembourg, L-1511 Luxembourg, Luxembourg}
\affiliation{Department of Physics and Astronomy, University of Pennsylvania, Philadelphia PA 19104}
\author{Charles L. Kane}
\affiliation{Department of Physics and Astronomy, University of Pennsylvania, Philadelphia PA 19104}
\date{\today}

\begin{abstract}
We develop the theory of an Andreev junction, which provides a method to probe the intrinsic topology of the Fermi sea of a two-dimensional electron gas (2DEG).   An Andreev junction is a Josephson $\pi$ junction proximitizing a ballistic 2DEG, and exhibits low-energy Andreev bound states that propagate \textit{along} the junction.   It has been shown that measuring the nonlocal Landauer conductance due to these Andreev modes in a narrow linear junction leads to a topological Andreev rectification (TAR) effect characterized by a quantized conductance that is sensitive to the Euler characteristic $\chi_F$ of the 2DEG Fermi sea.   Here we expand on that analysis and consider more realistic device geometries that go beyond the narrow linear junction and fully adiabatic limits considered earlier.   Wider junctions exhibit additional Andreev modes that contribute to the transport and degrade the quantization of the conductance.  Nonetheless, we show that an appropriately defined {\it rectified conductance} remains robustly quantized provided large momentum scattering is suppressed.   We verify and demonstrate these predictions by performing extensive numerical simulations of realistic device geometries.   We introduce a simple model system that demonstrates the robustness of the rectified conductance for wide linear junctions as well as point contacts, even when the nonlocal conductance is not quantized.  Motivated by recent experimental advances, we model devices in specific materials, including InAs quantum wells, as well as monolayer and bilayer graphene.  These studies indicate that for sufficiently ballistic samples observation of the TAR effect should be within experimental reach.

\end{abstract}

\maketitle

\section{Introduction} \label{sec:intro}

A powerful method for characterizing quantum many-body phases of matter is to identify quantized response functions that probe topological features of the phase.  This type of analysis was initiated by the integer quantized Hall effect, which probes the Chern number characterizing the topology of a gapped two-dimensional (2D) electronic phase \cite{Klitzing1980, TKNN1982}.   A related type of quantized response occurs in a one-dimensional (1D) ballistic metal, where the Landauer conductance exhibits steps that are 
quantized in units of $e^2/h$ reflecting the number of occupied one-dimensional subbands \cite{Landauer1957,FisherLee1981,Buttiker1986}. This integer can be interpreted as the Euler characteristic of the Fermi sea, $\chi_F$, a topological invariant that counts the number of components of the filled Fermi sea of a one-dimensional metal. Though the quantized Landauer conductance is insensitive to weak electron-electron interactions, this type of quantized response is less robust than the quantized Hall effect because it requires the conducting channels to be perfectly transmitted and the reflected channels to be perfectly reflected.   This will be approximately the case provided transport is sufficiently ballistic, the contacts are sufficiently reflectionless and the channel is sufficiently long.   Despite these conditions, quantized Landauer transport has been observed in a variety of systems, including relatively short quantum point contacts \cite{vanWees1988, Honda1995, Frank1998, vanWeperen2013}.

The feasibility of quantized Landauer transport in one dimension motivated the search for probes of Fermi sea topology in higher-dimensional metals.   Frequency-dependent and time-domain nonlinear responses \cite{Kane2022a,Yang2022,Zhang2022}, as well as equal-time density correlations \cite{Tam2022a}, have been shown to probe $\chi_F$, though unlike the Landauer conductance in one dimension, quantization of these quantities is only precise in the absence of interactions.   It was argued that a measure of multipartite entanglement in an interacting Fermi liquid probes $\chi_F$ \cite{Tam2022a}, but that is challenging to probe experimentally.

\begin{figure}
    \centering
    \includegraphics[width=0.95\linewidth]{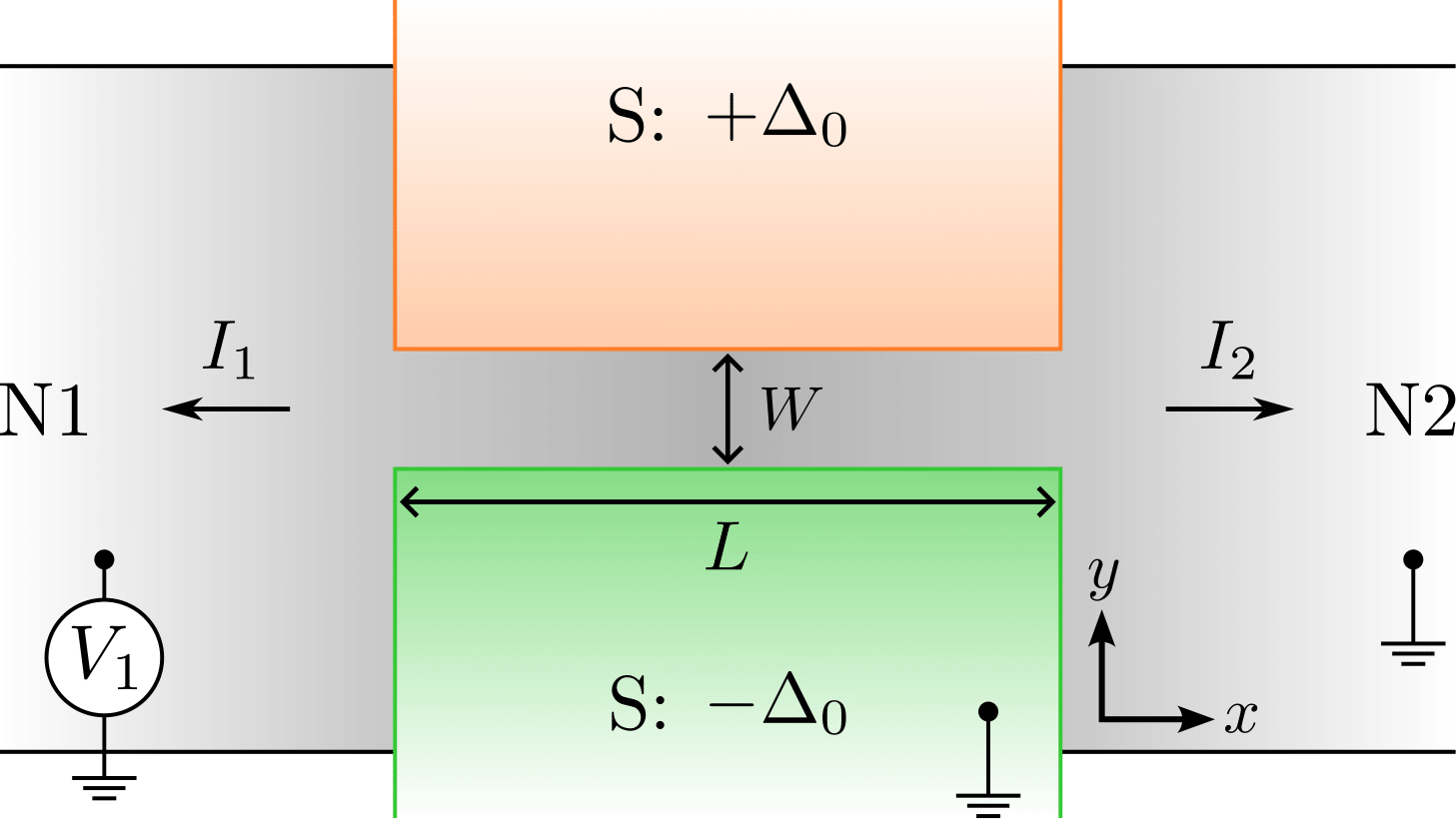}
    \caption{Layout of the Andreev junction, with width $W$ and length $L$.   A 2DEG is proximitized by grounded superconductors with a $\pi$ phase difference.   The topology and geometry of the Fermi sea of the 2DEG is probed by measuring the nonlocal conductance $G_{21} \equiv dI_2/dV_1$ between the normal metallic leads (N1 at bias $V_1$ and N2 grounded).}
    \label{fig:system}
\end{figure}

Recently, two of the authors introduced a method to probe $\chi_F$ of a two-dimensional electron gas (2DEG) that is robust in the presence of interactions \cite{Tam2022b}.
The proposal involves proximitizing the 2DEG with a grounded superconductor to form a linear\footnote{This type of device architecture has also been called \emph{planar} in the literature \cite{Pientka2017, Ren2019, Fornieri2019, Banerjee2022_1, Banerjee2022_2, Banerjee2022_3}. Here we use the term \emph{linear} to highlight the one-dimensional channel in the 2DEG between the superconductors. We will also consider a point-contact geometry, where the 1D channel is reduced to a point.} SNS (S: superconductor; N: normal metal) Josephson junction with phase difference $\pi$.  Then $\chi_F$ is probed by measuring the Landauer transport {\it along} the junction between source and drain contacts to the 2DEG at either end of the superconducting junction, as shown in Fig.\ \ref{fig:system}.   The current in the drain ($I_2$) in response to a voltage applied to the source ($V_1$) is carried by Andreev bound states (ABSs) that are extended along the junction.   We will refer to this setup as an {\it Andreev junction}, so as to contrast with the more conventional Josephson junction where the current of interest flows between the superconductors.    

It is instructive to compare transport in this Andreev junction setup to Landauer transport in an ordinary point contact or quantum wire.   In the case of ordinary Landauer transport, the 1D channel is formed by pinching off a 2DEG using an insulating energy gap. For the Andreev junction, the 1D channel is formed by pinching off with a {\it superconducting} energy gap.   For a wide junction, relative to the superconducting coherence length, there will be many channels of ABSs within the junction.   For a narrow junction, provided the phase difference across the junction is $\pi$, there will remain at least a single pair of ABSs that disperse as a function of the momentum along the junction.

In Ref.\ \cite{Tam2022b}, it was found that the source to drain Landauer conductance of an \textit{adiabatic} and \textit{narrow} Andreev junction is
given by
\begin{equation}\label{eq:keyresult0}
G_{21}(V) \equiv \left.\frac{dI_2}{dV_1} \right|_{V_1=V} = 
\begin{cases}
c_e \, e^2/h & \;\text{for}\;\; V< 0, \\
c_h \, e^2/h & \; \text{for}\;\; V > 0.
\end{cases}
\end{equation}
Here $c_e$($c_h$) are the number of propagating Andreev modes in the 1D channel that become electrons (holes) upon adiabatic evolution into the leads.   For a narrow linear junction, these, in turn, are determined by the number of electronlike (holelike) critical points on the Fermi surface, where the Fermi velocity is parallel to the channel, and the Fermi surface is locally convex (concave).   A novel feature of this result is that the conductance is different at positive and negative bias voltage.   This leads to a {\it rectification effect}, in which a low-frequency AC voltage in the source, $ V_{\text{AC}} =  V_0 \cos(\omega t)$, leads to a DC current flowing into the drain, $I_{\text{DC}} = -(e^2 |V_0|/\pi h)(c_e-c_h)$.   While the integers $c_e$ and $c_h$ depend on the specific geometry of the Fermi surface, as well as its orientation relative to the 1D channel, their difference depends only on its topology:
\begin{equation} \label{eq:cech}
c_e - c_h = \chi_F,
\end{equation}
where $\chi_F$ is the Euler characteristic of the 2D Fermi sea.   
In general, $\chi_F$ for any 2D Fermi sea can be expressed as the sum over all disconnected components of the Fermi surface, where an electronlike (holelike) Fermi surface contributes $+1$ ($-1$) and an open Fermi surface contributes $0$.
We will henceforth refer to this effect as {\it topological Andreev rectification} (TAR).  It motivates us to define the {\it rectified conductance}
\begin{equation} \label{eq:def_deltaG}
\delta G_{21}(V) \equiv G_{21}(V) - G_{21}(-V). 
\end{equation}
It follows from Eq.\ \eqref{eq:keyresult0} and Eq.\ \eqref{eq:cech} that $\delta G_{21}(V)$ probes the intrinsic topology of the 2D Fermi sea.

The Andreev junction resembles similar Josephson junction devices that have been studied in an effort to engineer proximity-induced topological superconductivity \cite{FuKane2008,Wieder2014,Hell2017, Pientka2017, Ren2019, Fornieri2019}.  Such systems based on InAs or HgTe quantum wells \cite{Hell2017, Pientka2017, Ren2019, Fornieri2019} are among the promising venues for studying the TAR effect.   Nonlocal transport measurements have proven to be a useful diagnostic for probing topological superconductivity in Josephson junction setups \cite{Banerjee2022_1,Banerjee2022_2,Banerjee2022_3}, as well as in proximitized quantum wires \cite{Danon2020,Menard2020}.   In that context, a rectification effect has been studied \cite{Rosdahl2018}, which results from the specific structure of the Andreev levels in a proximitized 1D nanowire. Rectification in Andreev interferometers that require broken time-reversal symmetry has also been studied \cite{Meair2012}. However, these are all distinct from the TAR effect, which probes the intrinsic topology of the Fermi sea of a 2DEG.

In this paper, we will expand on the analysis in Ref. \cite{Tam2022b}, and in an effort to hasten experimental demonstration of the TAR effect, we will consider more realistic device geometries that go beyond the narrow junction and fully adiabatic limits considered in Ref.\ \cite{Tam2022b}.   We will begin with a review of the narrow junction limit in Section \ref{sec:review}.   Wider junctions will be considered in Section \ref{sec:smatrix}.   In the latter case there exist additional Andreev modes bound to the SNS junction.   These modes will be critical to our analysis because they can contribute to the nonlocal conductance, and when there is scattering among them, the quantization is degraded. We will identify a physically accessible regime in which interchannel scattering among modes with nearly the same momentum, that are associated with the same Fermi surface critical point, degrades the quantization of the nonlocal conductance $G_{21}$.   However, we will show by analyzing the multichannel Landauer transmission problem that, provided large momentum scattering {\it between} Fermi surface critical points is absent, the rectified conductance $\delta G_{21}$ remains quantized.  Thus, the rectified conductance provides a more robust marker for the intrinsic topology of the Fermi sea.

In order to verify these predictions, we have performed extensive simulations of realistic device geometries using the \textsc{Kwant} package \cite{Groth2014, codes}.   In Section \ref{sec:model} we introduce a simple toy model that allows us to consider a wide variety of Fermi sea topologies and real-space junction geometries that access several regimes of interest, including narrow junctions, wide junctions, as well as a point-contact geometry.   We find that in all of these cases, the TAR effect can be observed provided the transport is sufficiently ballistic.
In Section \ref{sec:materials}, we model junctions in specific real materials that are motivated by recent experimental studies of Josephson junctions in InAs quantum wells, as well as in monolayer and bilayer graphene.   In these simulations we assess the feasibility of measuring the TAR effect using experimental parameters.

\section{Topological Andreev rectification} \label{sec:TAR}

In this section we develop the analytic framework for topological Andreev rectification.   We will begin in Section \ref{sec:review} by reviewing the analysis in Ref.\ \cite{Tam2022b} of the fully adiabatic narrow-junction limit.   In Section \ref{sec:smatrix} we  introduce a multichannel scattering analysis that allows for the analysis of wider junctions that are not perfectly adiabatic.   We will show that provided large momentum scattering is suppressed, the rectified conductance $\delta G_{21}$ remains a quantized probe of the Fermi sea topology, even when the quantization of the nonlocal conductance $G_{21}$ is degraded.

\subsection{Review} \label{sec:review}

It was established in Ref.\ \cite{Tam2022b} that the rectified nonlocal conductance of a narrow and adiabatic Andreev junction is quantized and probes the Euler characteristic of the 2D Fermi sea.   Here we review that analysis. 

\subsubsection{Euler characteristic}

\begin{figure}
    \centering
    \includegraphics[width=0.6\linewidth]{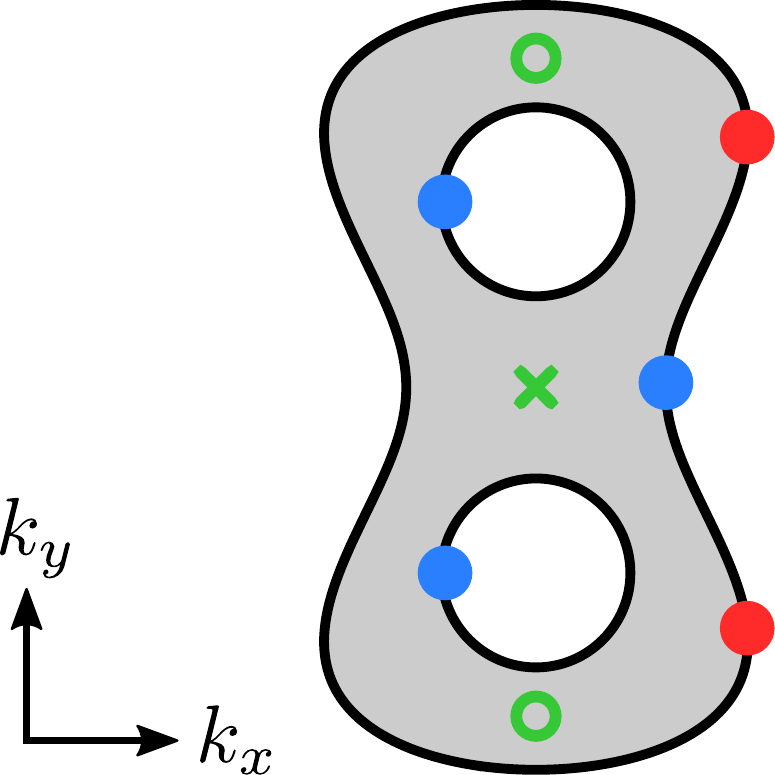}
    \caption{Illustration  of a Fermi sea (gray region) in two spatial dimensions. Fermi sea critical points are shown as green crosses (minima) and circles (saddles). Fermi surface (black line) convex/concave critical points, defined for $+\hat{x}$, are indicated by red/blue dots with their count denoted as $c_{e/h}$. Here $(c_e, c_h) = (2,3)$, giving $\chi_F = c_e - c_h = -1$.}
    \label{fig:fermi}
\end{figure}

The Euler characteristic is an integer topological invariant that characterizes any topological space.   Here we review several equivalent formulations of the Euler characteristic, $\chi_F$, of a two-dimensional Fermi sea.  For concreteness, consider the hypothetical Fermi sea with $\chi_F=-1$ shown in Fig.\ \ref{fig:fermi}.

{\it a. In terms of Betti numbers}:   The Euler characteristic may be written as \cite{nakahara2003geometry, Dieck2008}
\begin{equation}
    \chi_F = \sum_l (-1)^l b_l,
\end{equation}
where the Betti numbers $b_\ell$ count the number of topologically distinct $\ell$-cycles.   In Fig.\ \ref{fig:fermi}, the Fermi sea has a single connected component, hence $b_0=1$. The two holes in the Fermi sea imply two independent $1$-cycles, hence $b_1=2$, and $b_\ell=0$ for $\ell>1$.

{\it b. In terms of critical points of $E({\bm k})$}:  According to Morse theory, $\chi_F$ can be expressed in terms of the critical points of a Morse function \cite{Milnor1963, Nash1988}.   The electronic dispersion $E({\bm k})$ is a natural Morse function, and the critical points where $\nabla_{\bm k} E({\bm k})=0$  are characterized by a Morse index $\eta = (0,1,2)$ for a (minimum, saddle, maximum).  Then,
\begin{equation}
    \chi_F = n_0 - n_1 +n_2,
\end{equation}
where $n_\eta$ is the number of critical points inside the Fermi sea with index $\eta$.   Figure \ref{fig:fermi} identifies one minimum $(\eta=0)$ and two saddles ($\eta=1$).  This formulation makes it clear that a topological Lifshitz transition occurs when a Fermi sea critical point passes through $E_F$ \cite{Lifshitz1960, Volovik2017}.

{\it c. In terms of genus and boundary components:}   In two dimensions $\chi_F$ can be expressed as a sum over all connected components $k$ of the Fermi sea,
\begin{equation}
\chi_F = \sum_k (2 - 2g_k - b_k),
\end{equation}
where $g_k$ is the genus of component $k$ and $b_k$ is the number of boundary components.   In Fig.\ \ref{fig:fermi}, there is a single component with $g=0$ and $b=3$, so $\chi_F = -1$.  A non-trivial genus can arise because the Brillouin zone is a torus.   For example, the complement of the Fermi sea in Fig.\ \ref{fig:fermi} has three components.  Each of the two holes has $g=0$ and $b=1$, while the outer component wraps the Brillouin zone and thus has $g=1$ and $b=1$, leading to $\chi_F=+1$ altogether.  In general, $\chi_F$ is odd under the exchange of electrons and holes for an even-dimensional Fermi sea. 

{\it d. In terms of the Fermi surface:}   In two dimensions every component of the Fermi surface has the topology of a circle, but they can be distinguished by whether the circle encloses electrons, encloses holes, or is open (i.e., encloses neither electrons nor holes). Then,
\begin{equation}
    \chi_F = n_e - n_h,
\end{equation}
where $n_{e(h)}$ counts the number of electronlike (holelike) Fermi surfaces, whereas open Fermi surfaces contribute zero. In Fig.\ \ref{fig:fermi}, we have $n_e=1$ and $n_h=2$.

{\it e. In terms of Fermi surface critical points:}  If we specify an arbitrary unit vector $\hat \zeta$, then we may identify critical points on the Fermi surface to be the points where the velocity $\bm v(\bm k) = \hbar^{-1} \nabla_{\bm k}E({\bm k}) \parallel \hat \zeta$.   Fermi surface critical points can be distinguished by the sign of the curvature of the Fermi surface at that point.   We call the critical point {\it convex}, or {\it electronlike}, if $\partial^2E/\partial k_\perp^2 >0$ (where $k_\perp$ is the component of ${\bm k}$ perpendicular to $\hat \zeta\,$).   In the vicinity of such a critical point the Fermi surface resembles an electron pocket.   Alternatively, the critical point is {\it concave}, or {\it holelike}, if $\partial^2E/\partial k_\perp^2 <0$, and the Fermi surface locally resembles a hole pocket.   In terms of these data,
\begin{equation}
    \chi_F= c_e - c_h,
\end{equation}
where $c_{e(h)}$ gives the number of electronlike (holelike) critical points.   In Fig.\ \ref{fig:fermi}, the electronlike and holelike critical points for $\hat \zeta = +\hat x$ are, respectively, shown as the red and blue dots, with $c_e=2$ and $c_h=3$.

For the purposes of this paper this last formulation provides the most natural connection to the Andreev junction.   We will show that the low-energy transport in the junction is dominated by the Fermi surface critical points where the velocity is parallel to the junction (i.e., with $\hat \zeta$ parallel to the NS interfaces).   It is worth emphasizing that the integers $c_e$ and $c_h$ are not by themselves topological invariants.   They depend on the specific geometry of the Fermi surface as well as on the chosen direction for defining the critical points.   However, the difference $c_e-c_h$ depends only on the intrinsic topology of the Fermi sea.   For example, if instead in Fig.\ \ref{fig:fermi} we had chosen $\hat \zeta = +\hat y$, then we would have $c_e = 1$ and $c_h=2$.  

\subsubsection{Andreev zero modes at a $\pi$ junction}

Fermi surface critical points, as defined above, are probed by Andreev bound states (ABSs). In this work, as well as in Ref.\ \cite{Tam2022b}, we focus on the ABSs hosted by an SNS $\pi$ junction (i.e., with a superconducting phase difference of $\pi$, see Fig.\ \ref{fig:system}). The reason for focusing on $\pi$ junctions is that in this case the presence of subgap ABSs is a robust topological feature.   To see this, consider an infinitely-long channel oriented in the $\hat x$ direction, such that the electronic states can be indexed by the momentum $k_x$.  The Andreev states bound to the junction are found by solving the Bogoliubov-de Gennes (BdG) equation for motion in the $\hat y$ direction for each value of $k_x$.   
For values of $k_x$ that are away from the Fermi surface critical points (defined for $\hat \zeta = \pm \hat x$) the electronic dispersion along the $k_y$ direction can be approximated in the vicinity of a Fermi surface point $k_F^y(k_x)$ by linearizing: $E(k_x,k_y) \approx E_F +  v_y(k_x) \left[ k_y - k_F^y(k_x) \right]$ (here and throughout this section we set $\hbar=1$).   When there is a single pair of Fermi surface points at $ \bm k = (k_x, \pm k_F^y)$ with opposite velocity $\pm v_y$, the resulting BdG equation takes the form of a 1D Dirac equation with a spatially-varying mass term due to the pairing potential $\Delta= \Delta_1+i\Delta_2$. The corresponding BdG Hamiltonian is
\begin{equation}
    \mathcal H = -i \tau_z \sigma_z v_y \partial_y  + \Delta_1(y) \tau_x + \Delta_2(y) \tau_y,
    \label{jackiw-rebbi}
\end{equation}
where $\sigma_z = \pm 1$ distinguishes the Fermi points, and $\bm \tau$ are Pauli matrices in the Nambu space.   
When the phase difference across the junction is $\pi$, we may set $\Delta_1=0$, and $\Delta_2(y)$ changes sign across the junction. Equation \eqref{jackiw-rebbi} then  exhibits a pair of degenerate Jackiw-Rebbi zero modes indexed by $\sigma_z$ \cite{Jackiw1976}.    For values of $k_x$ where there are no Fermi surface points, then clearly there will be no zero modes.   If $k_x$ is such that there are $2n$ Fermi surface points, then there will be $2n$ zero modes.   Thus, the count of zero modes of Eq.\ \eqref{jackiw-rebbi} as a function of $k_x$ contains geometric information about the Fermi sea.

The exactness of the zero-mode solutions of Eq.\ \eqref{jackiw-rebbi} relies on the absence of any normal reflection at the interface, which guarantees that $[\mathcal H,\sigma_z]=0$.   In this work, we focus on junctions that are perfectly transparent, so that the Hamiltonian contains no scattering terms that connect the Fermi points.   However, even for a perfect junction, the exactness of the zero modes is an artifact of the linear approximation in Eq.\ \eqref{jackiw-rebbi}, which treats the $\sigma_z=\pm 1$ bands as completely independent.   In reality, states at different Fermi points can derive from the same band, and a more complete description can be formulated in terms of a single-band model described by a second (or higher) order differential equation.   As shown below, this leads to a splitting of the zero modes by an amount of order $|\Delta|^2/\mu(k_x)$, where $\mu(k_x)$ is the Fermi energy relative to the band edge at a given $k_x$.   

When $k_x$ approaches a Fermi surface critical point, the linearized approximation necessarily breaks down because $\mu(k_x) \rightarrow 0$.  This is consistent with the fact that when $k_x$ crosses a critical point the zero mode count will change.   In the vicinity of the critical points the energy of the nominal zero modes will disperse as a function of $k_x$.   Our working assumption in this paper is that $|\Delta| \ll E_F$, where $E_F$ is the appropriate energy scale associated with the Fermi sea. Deep inside the Fermi sea (away from the surface critical points), the splitting of the zero modes is thus of order $|\Delta|^2/E_F$, which is much smaller than $|\Delta|$.   In this case, the zero modes will exhibit significant dispersion as a function of $k_x$ \emph{only} near the critical points.   Thus, the dispersing Andreev modes are in correspondence with the critical points, and can be analyzed by focusing on the vicinity of each critical point.   This analysis was introduced in Ref.\ \cite{Tam2022b}, and is reviewed below.

\subsubsection{Dispersive Andreev modes associated with Fermi surface critical points}

Close to a Fermi surface critical point at which $v_y(k_x) = \partial E/\partial k_y = 0$, the energy can be linearized as a function of $k_x$, but it is necessary to expand to second order in $k_y$.   This leads to the following model for an Andreev junction of width $W$ \cite{Tam2022b}, 
\begin{equation}\label{eq:HBdG_criticalpt}
\begin{split}
    H_{\text{BdG}} (k_x) = \; & \left[ \theta \left( y - W/2 \right) - \theta \left( -y - W/2 \right) \right] \Delta_0 \tau_y \\
    & + \left( -\frac{1}{2m^*} \, \partial^2_y + v_x\delta k_x \right) \tau_z.
\end{split}
\end{equation}
Here $\delta k_x \equiv k_x - k_F$ is the $k_x$ momentum relative to the critical point at $\bm k = (k_F, 0)$, $\tau_y$ and $\tau_z$ are Pauli matrices acting in the electron-hole space, $\Delta_0$ is the magnitude of the proximity-induced pairing gap, and $m^* \gtrless 0$ is the effective mass at the convex/concave critical point. There are two symmetries useful for finding ABSs: (1) Chiral symmetry $\tau_x$, which relates states of energy $\pm \varepsilon$ for a fixed $\delta k_x$. (2) Mirror symmetry $ \tau_z \mathcal{M}_y$ (where $\mathcal{M}_y$ takes $y \mapsto -y$), which labels each ABS by a mirror eigenvalue $\pm 1$. Clearly, electron-hole partners (related by $\tau_x$) acquire opposite mirror eigenvalues.

\begin{figure}
    \centering
    \includegraphics[width=\linewidth]{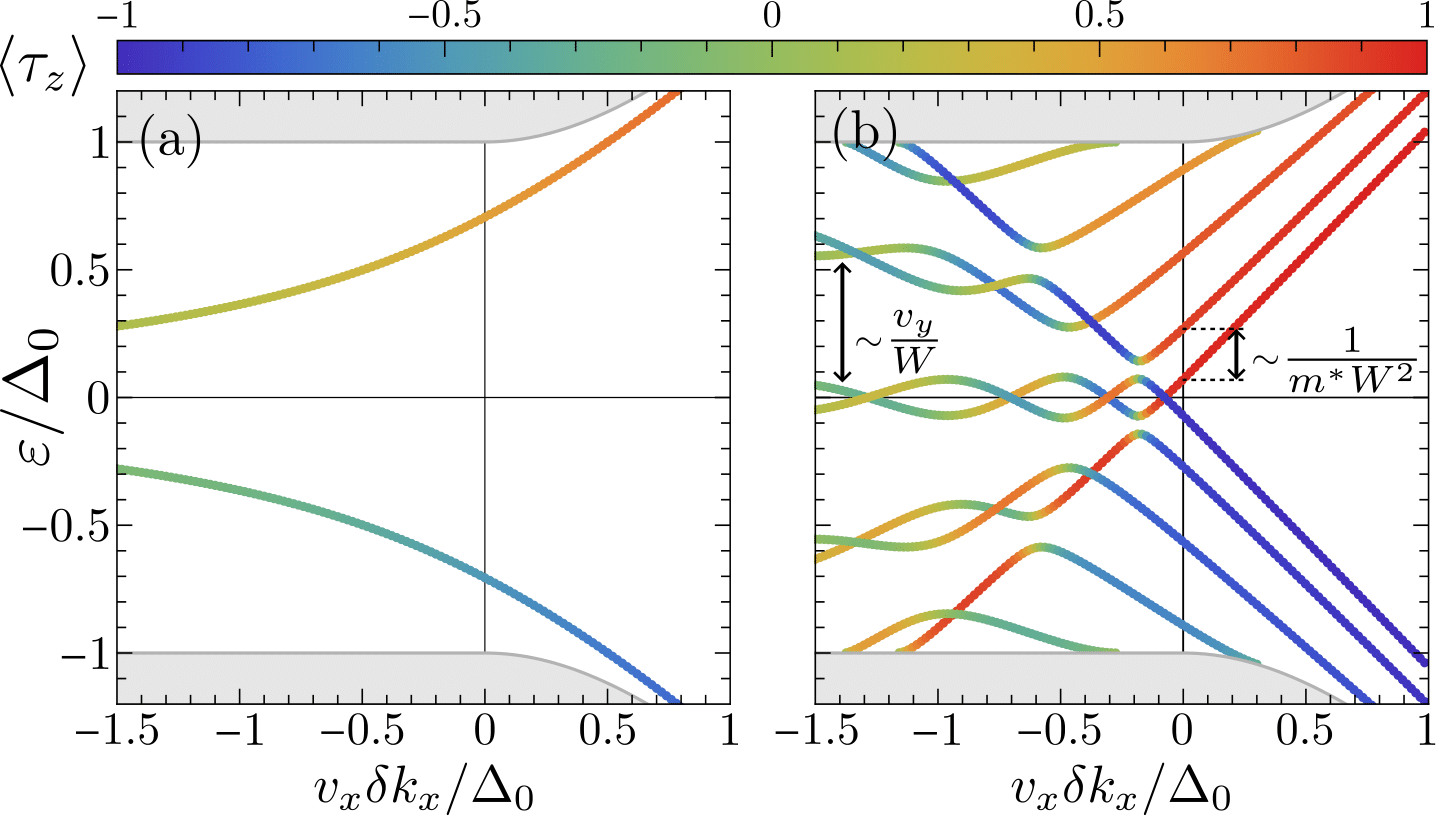}
    \caption{ABS spectrum of an infinitely-long $\pi$ junction near a convex Fermi surface critical point. Shown here for (a) a narrow junction with $W \sqrt{|2m^*\Delta_0|} = 0.1$ and (b) a wide junction with $W \sqrt{|2m^*\Delta_0|} = 10$. In (b) we indicate the scaling of the level spacing with $W$ near the critical point ($\sim 1/m^*W^2$) and deep in the Fermi pocket ($\sim v_y/W$) where $v_y(k_x)$ is the Fermi velocity in the $k_y$ direction. The crossings are protected by the mirror symmetry $\tau_z \mathcal M_y$ and the color scale gives the electron-hole character $\left< \tau_z \right> = v_x^{-1} d\varepsilon / dk_x$ of the ABSs.}
    \label{fig:critical-spectrum}
\end{figure}

For $W=0$ exactly one pair of ABSs is found with dispersion \cite{Tam2022b}
\begin{equation} \label{eq:dispersion}
    \varepsilon (k_x) = \pm \left[ \frac{ v_x \delta k_x}{2} + \sgn(m^*) \sqrt{\left( \frac{v_x\delta k_x}{2} \right)^2+\frac{\Delta_0^2}{2}} \, \right].
\end{equation} 
Deep inside the Fermi pocket where $v_x \delta k_x / m^* < 0 $, the dispersion of the ABS is flattened with $\varepsilon \approx \pm\Delta_0^2 / ( 2 v_x \delta k_x )$.  For $\delta k_x \sim k_F$, i.e., the momentum scale associated with the size of the Fermi sea, we have $v_x \delta k_x \sim E_F$ which recovers the $\Delta_0^2/E_F$ splitting mentioned above, and is consistent with the expectation of having approximate zero modes for $\Delta_0\ll E_F$.   On the other hand, far outside the Fermi pocket, where $v_x \delta k_x / m^* > 0$, the ABS energy approaches the dispersion of the metal with $\varepsilon \approx \pm v_x \delta k_x$, so the zero mode is absent.   Right at the critical point $\varepsilon = \pm \Delta_0/\sqrt{2}$.
While the exact form of the dispersion depends on microscopic details (e.g., $v_x$ and $W$), the presence of such a dispersive ABS for each Fermi surface critical point is a robust topological feature of the narrow $\pi$ junction. Figure \ref{fig:critical-spectrum}(a) shows the spectrum of the ABS near a convex critical point for a finite-width $\pi$ junction with $W \ll |2m^*\Delta_0|^{-1/2}$, which is well approximated by \eqref{eq:dispersion}.

Just as there are two types of Fermi surface critical points (convex or concave), there are also two types of dispersive ABSs, which are distinguished by whether they are electronlike or holelike at positive energy.   We define the electron-hole character of an ABS excitation as $\langle\tau_z \rangle$ for the BdG mode with $\varepsilon >0$.  This is equal to $+1$ ($-1$) for a normal electron (hole).   Note that the electron-hole character is related to the slope of the ABS dispersion by the Hellman-Feynman theorem:  $d\varepsilon/dk_x = v_x \langle\tau_z \rangle$.  Thus, the electron-hole character, indicated by the color of the curves in Fig.\ \ref{fig:critical-spectrum}, is related to the slope of the curves. As we have established that the ABS dispersion $\varepsilon \rightarrow \pm  v_x\delta k_x$ far outside the Fermi pocket, we find $\langle \tau_z \rangle \rightarrow \pm 1$.  A convex (electronlike) critical point [as shown in Fig.\ \ref{fig:critical-spectrum}(a)] is associated to an electronlike ABS excitation with a positive electron-hole character.   A concave (holelike) critical point will have a dispersion that resembles Fig.\ \ref{fig:critical-spectrum}(a) rotated by $180^\circ$, and will be associated with a holelike ABS excitation with a negative electron-hole character.

The electron-hole character of an ABS excitation, in turn, determines the fate of the mode as it propagates into a normal lead.  In Ref.\ \cite{Tam2022b}, a lead was modeled in two ways.   The simplest is to imagine adiabatically decreasing $\Delta_0$ (while keeping $W$ fixed).   In that case, the energy of the single ABS excitation relative to $\Delta_0$ grows, so that in Fig.\ \ref{fig:critical-spectrum}(a) $\varepsilon/\Delta_0$ moves away from zero.  An initially low-energy Andreev state, with a mixed electron-hole character, will thus evolve into a state with a \textit{definite} electron-hole character $\langle\tau_z\rangle = \pm 1$.   A somewhat more realistic model for an adiabatic contact is to consider increasing $W$ sufficiently slow (keeping $\Delta_0$ fixed).  As $W$ increases, there will be additional Andreev modes bound to the junction, as shown in Fig.\ \ref{fig:critical-spectrum}(b).   In the adiabatic approximation, a state will follow the same band as it smoothly evolves upon increasing $W$, and it can again be seen that deep inside the lead an ABS excitation will acquire a definite electron or hole character, which is determined by whether the critical point is convex or concave.   This, in turn, determines the current carried by a quasiparticle populating the Andreev mode when it propagates into a lead.   It also specifies the population of the Andreev mode, which depends on the chemical potential of the lead it is incident from.

We thus conclude that each electronlike (holelike) critical point is associated with a propagating ABS that provides a conducting channel connecting the source and the drain for electrons (holes).  We next show that these modes lead to a quantized Landauer conductance that depends on the sign of the voltage bias.

\subsubsection{Quantized transport via Andreev states}

We now apply the relation between the convex or concave nature of a Fermi surface critical point and the electron or hole nature of the associated dispersive ABS to transport along a narrow Andreev junction as depicted in Fig.\ \ref{fig:system}. We are interested in the nonlocal differential conductance defined as (see Appendix \ref{app:landauer}),
\begin{equation}
    G_{21} \equiv \frac{dI_2}{dV_1} \overset{T \rightarrow 0}{=} \frac{e^2}{h} \left[ T^{ee}_{21}(\varepsilon) - T^{he}_{21}(\varepsilon) \right]_{\varepsilon = eV_1}.
\end{equation}
Here $T^{ee(he)}_{21}(\varepsilon)$ describes the transmission of an electron incident from lead N1 at energy $\varepsilon$ to lead N2, where it emerges as an electron (hole).   For $\varepsilon<0$ this could equally well be interpreted as $T^{hh(eh)}_{21}(-\varepsilon)$, i.e., the transmission of a hole at energy $-\varepsilon$ in N1 to a hole (electron) in N2.
The equivalence of these interpretations is guaranteed by the electron-hole symmetry of the BdG theory.
When the junction length $L$ is much longer than the superconducting coherence length
$\xi = v_x / \pi \Delta_0$, crossed Andreev reflection is suppressed and hence $T^{he}_{21}=0$. Moreover, the dispersive ABSs are the only transmission channels between leads N1 and N2. Since there are $c_{e}$ ($c_h$) number of electronlike (holelike) ABSs, and assuming reflectionless contacts and ballistic transport, we find
\begin{equation}
    T^{ee}_{21}(\varepsilon>0) = c_e \quad \text{and}\quad T^{ee}_{21}(\varepsilon<0) = c_h, 
\end{equation}
which implies
\begin{equation}\label{eq:keyresult1}
    G_{21} = \frac{e^2}{h} \left[ c_e \, \theta(eV_1) + c_h \, \theta(-eV_1) \right].
\end{equation}
Notice that the quantization in $G_{21}$ is sensitive to the shape of the Fermi sea as well as the orientation of the junction. On the other hand, the rectified nonlocal conductance, as defined in Eq.\ \eqref{eq:def_deltaG}, obeys
\begin{equation}\label{eq:keyresult2}
    \delta G_{21}(V_1) = \sgn(eV_1) \, \frac{e^2}{h} \, \chi_F, 
\end{equation}
which is a quantized probe of the intrinsic topology of the Fermi sea. We refer to the quantization in $\delta G_{21}$ as the topological Andreev rectification (TAR) effect. 

Equation \eqref{eq:keyresult1} is the central result of Ref.\ \cite{Tam2022b}, which is derived for a narrow Andreev junction hosting one ABS per Fermi surface critical point, and under the assumption that the ABSs moving out of the junction would adiabatically evolve into definite electrons or holes inside the normal lead. In this case, Eq.\ \eqref{eq:keyresult2} is simply a corollary.   However, as we argue below, for a wider junction, which can have multiple ABSs, there is a regime in which the quantization of $G_{21}$ is degraded, while the quantization of $\delta G_{21}$ remains robust.

The reason for the existence of this regime can be understood by examining the Andreev mode dispersion near a critical point for a wider junction, as shown in Fig.\ \ref{fig:critical-spectrum}(b). We first note that the ABSs of a wide Andreev junction tend to bundle in pairs with opposite mirror eigenvalues.  This is related to the oscillations in the spectrum which originate from anticrossings between the same $\tau_z \mathcal{M}_y$ mirror sector and symmetry-protected crossings between different mirror sectors. In particular, at zero energy, different sectors have to cross because $\tau_z \mathcal{M}_y$ anticommutes with chiral symmetry $\tau_x$ \cite{Tam2022b}. The zero-energy crossings are determined by $k_y^2/2m^* + v_x \delta k_x = 0$ where $k_y$ is obtained from the boundary conditions. In the limit $\Delta_0 / v_x \delta k_x \gg 1$, the wave function vanishes at $y=\pm W/2$. This gives the condition $k_y = n\pi/W$ ($n=1,2,\ldots$) and hence $v_x \delta k_x/\Delta_0 = - \sgn(m^*) ( n\pi / W \sqrt{|2m^*\Delta_0|} \, )^2$, which holds approximately in Fig.\ \ref{fig:critical-spectrum}(b). We further note that the Andreev level spacing is smaller when $k_x$ is close to the critical point.   In general, the Andreev level spacing is of order $v_y(k_x)/W$. In Eq.\ \eqref{eq:HBdG_criticalpt}, $v_y = k_F^y(k_x)/m^* \sim \left| v_x \delta k_x / m^* \right|^{1/2}$.   Deep inside the Fermi pocket, $\delta k_x \sim k_F$ and $v_y \sim v_F$, so the deep Andreev level spacing is of order $v_F/W$.   However, $v_y \rightarrow 0$ when $\delta k_x$ is small. This leads to a level spacing of order $1 / m^* W^2$ close to the critical point.    We will focus on the regime in which the energy of the relevant dispersive Andreev modes, which is set by the bias voltage $V_1$, is of order (or less than) the deep level spacing $v_F/W$.   In this case there can still be several Andreev bound states for $k_x$ close to the critical point.   These modes have nearly the same momentum, which makes the adiabatic approximation less reliable for them.   Moreover, even when the width of the junction increases adiabatically slow, the Fermi energy will pass through turning points where these modes come and go in a complicated way.   This inevitably leads to a breakdown in the adiabatic approximation, resulting in a degradation of the quantization of the nonlocal conductance $G_{21}$. In the following, we will show that despite this breakdown in the quantization of $G_{21}$, the rectified conductance $\delta G_{21}$ remains robustly quantized \emph{provided} large momentum scattering between the Fermi surface critical points is negligible.

\subsection{Beyond the narrow-junction limit} \label{sec:smatrix}

In this section we consider a junction of width $W \gtrsim |2 m^* \Delta_0|^{-1/2}$, which can have multiple ABSs associated with each Fermi surface critical point.   We analyze the coupling of these modes to the leads within a multichannel Landauer-B\"uttiker formalism.  Remarkably, we find that the quantization of the \textit{rectified} nonlocal conductance $\delta G_{21}$ persists even for a wide junction as long as scattering between ABSs across the Fermi surface is suppressed, and either time-reversal symmetry $\mathcal T$ or mirror symmetry $\mathcal M_x$ ($x\mapsto-x$) along the transport direction is preserved. We further find that $\delta G_{21}$ is related to the number of occupied ABSs deep in the Fermi pockets. Hence, at sufficiently small bias voltage the plateau in $\delta G_{21}$ remains a good measure of the Fermi sea topology $\chi_F$. 

\begin{figure}
    \centering
    \includegraphics[width=\linewidth]{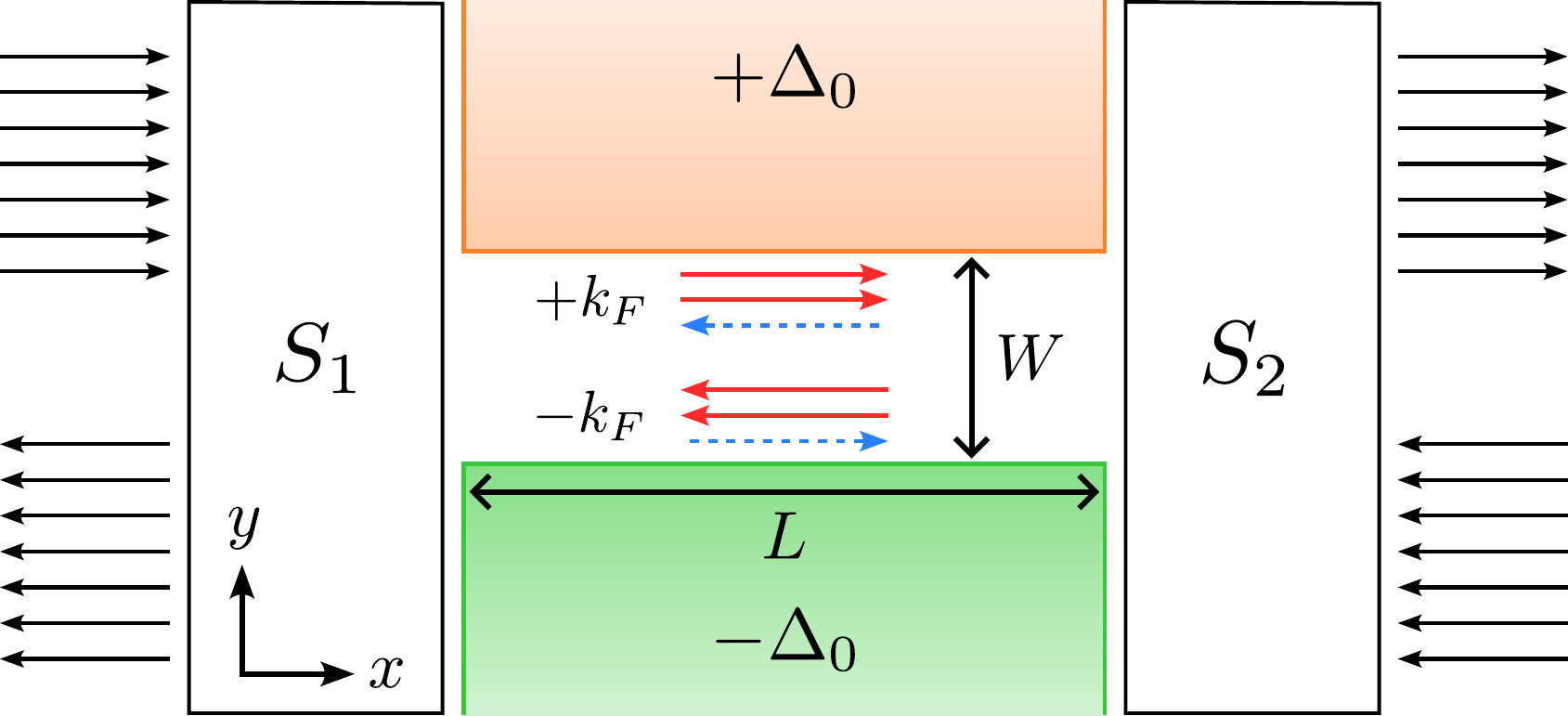}
    \caption{Model for transport along the Andreev junction. Scattering from the leads to the junction is described by the scattering matrices $S_1$ and $S_2$. Here we illustrate a case relevant for a convex Fermi surface critical point at $k_x=k_F$, where at a positive energy the number of electronlike (solid red) and holelike (dashed blue) modes in the junction at $\pm k_F$ are given by $N_e=2$ and $N_h=1$.}
    \label{fig:smatrix}
\end{figure}

We start by dividing the total system into the two leads and the junction, where the latter is defined as the region with constant width $W$. As illustrated in Fig.\ \ref{fig:smatrix}, scattering from the two leads to the junction is governed by two scattering matrices $S_1$ and $S_2$. We have
\begin{equation}
    \begin{pmatrix} b_{1L}^\text{lead} \\[.5mm] \phi_R^\text{ABS} \end{pmatrix} = S_1 \begin{pmatrix} a_{1R}^\text{lead} \\[.5mm] \phi_L^\text{ABS} \end{pmatrix}, \,\,
    \begin{pmatrix} b_{2R}^\text{lead} \\[.5mm] \phi_L^\text{ABS} \end{pmatrix} = S_2 \begin{pmatrix} a_{1L}^\text{lead} \\[.5mm] \phi_R^\text{ABS} \end{pmatrix},
\end{equation}
with $a$ and $b$ the incoming and outgoing amplitudes of propagating modes in the leads, and the subscripts indicating the lead and the propagation direction. The amplitudes of right- or left-propagating ABSs are given by $\phi_{R/L}^\text{ABS}$. Here we have absorbed the phases accumulated during propagation into the $S$ matrices,
\begin{equation}
    S_1 = \left( \begin{array}{c|c}
    r_{1L} & t_{1L} \\
    \hline
    t_{1R} & r_{1R}
    \end{array} \right), \qquad S_2 = \left( \begin{array}{c|c}
    r_{2R} & t_{2R} \\
   \hline
    t_{2L} & r_{2L}
    \end{array} \right),
\end{equation}
where the subscript $R/L$ denotes transmission ($t$) or reflection ($r$) to the right/left. 

\begin{figure}
    \centering
    \includegraphics[width=\linewidth]{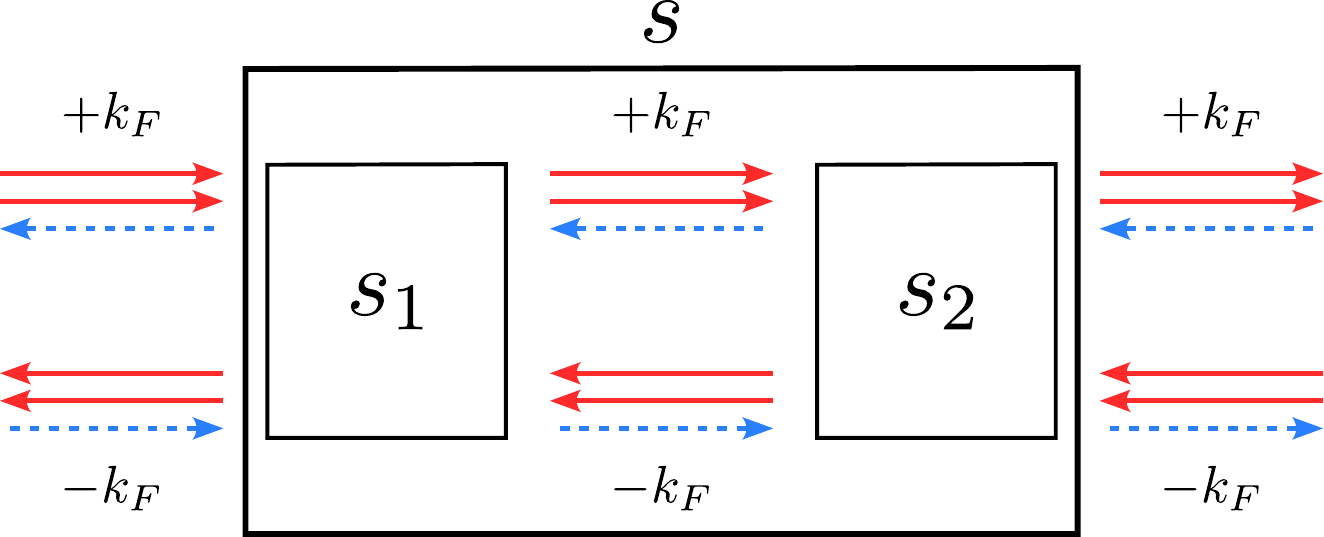}
    \caption{Decoupling of the reduced scattering problem into two independent channels at $\pm k_F$. Here, we illustrate the case with $N_e = 2$ electronlike modes (solid red) and $N_h = 1$ holelike modes (dashed blue) at positive energy, for a convex Fermi surface critical point at $k_x=k_F$. The total $S$ matrix in the nonzero transmission subspace is given by $s=s_1 \odot s_2$.}
    \label{fig:reduced-smatrix}
\end{figure}

Let us consider scattering between lead N1 and the junction in closer detail. The transmission matrix from lead N1 to the junction is the $N \times M_1$ matrix $t_{1R}$ where $N$ ($M_1$) is the number of right movers in the junction (lead N1). Since $M_1 \gg N$, the matrix $t_{1R}^\dag t_{1R}$ has at least $M_1-N$ zero eigenvalues by the rank-nullity theorem, which also holds for $t_{1L} t_{1L}^\dag$. We now choose a basis for the incoming (outgoing) lead modes that diagonalizes $t_{1R}^\dag t_{1R}$ (and $t_{1L} t_{1L}^\dag$). In this basis, we can write
\begin{equation}
    \begin{pmatrix} b_{1L}^\text{lead} \\ \phi_R^\text{ABS} \\[0.3mm] b_{1L}^\mathrm{\overline{lead}} \end{pmatrix} = S_1 \begin{pmatrix} a_{1R}^\text{lead} \\ \phi_L^\text{ABS} \\[0.3mm] a_{1R}^\mathrm{\overline{lead}} \end{pmatrix},   
\end{equation}
where the lead modes with zero transmission eigenvalues are denoted as $\mathrm{\overline{lead}}$ and
\begin{equation}
    S_1 = \left( \begin{array}{c|c|c}
    r_{1L} & t_{1L} & A \\
    \hline
    t_{1R} & r_{1R} & \bm 0_{N \times (M_1-N)} \\
    \hline
    B & \bm 0_{(M_1-N) \times N} & \rho_{1L}
    \end{array} \right).
\end{equation}
Here we have used the same notation even though the submatrices are generally different after the basis transformation. In the new notation $r_{1L}$, $r_{1R}$, $t_{1L}$, and $t_{1R}$ are $N\times N$ matrices\footnote{In the event that the original $t_{1R}^\dag t_{1R}$ (or $t_{1L} t_{1L}^\dag$) has more than $M_1-N$ zero modes, then there will exist zero modes of $t_{1R}t_{1R}^\dag$ (or $t_{1L}^\dag t_{1L}$), implying modes in the junction that are perfectly reflected and decoupled from the lead.  In that case the new $t_{1L}$ and $t_{1R}$ matrices are not square, which complicates our analysis. To resolve that, we add a small perturbation that lifts the extra zero modes, making $t_{1L}$ and $t_{1R}$ square.  That perturbation can then be set to zero in the decoupled scattering problem.}
, while $\rho_{1L}$ is square with dimension $M_1-N$. The latter contains both normal and Andreev reflection coefficients of the zero transmission lead subspace. Moreover, unitarity of the scattering matrix $S_1^\dag S_1 = S_1 S_1^\dag = \mathds 1$ implies
\begin{equation}
    t_{1L}^\dag A = t_{1R} B^\dag = 0.
\end{equation}
By construction,  $t_{1L}^\dag$ and $t_{1R}$ have trivial nullspaces.  It follows that $A$ and $B$ are zero. We obtain
\begin{equation} \label{eq:s1}
    S_1 = \left( \begin{array}{c|c|c}
    r_{1L} & t_{1L} & 0 \\
    \hline
    t_{1R} & r_{1R} & 0 \\
    \hline
    0 & 0 & \rho_{1L}
    \end{array} \right).
\end{equation}
Because we can perform the same procedure for the lead N2, the scattering problem is decoupled into three independent problems: two purely reflection problems of dimension $M_i-N$ ($i=1,2$ ) for the zero transmission lead modes and an $N$-dimensional problem involving both the junction and the lead modes with nonzero transmission eigenvalues. Since we are interested in the latter, we can henceforth focus on the smaller subsystem with $N \times N$ scattering matrices $s_1$ and $s_2$, as illustrated in Fig.\ \ref{fig:reduced-smatrix}.

Up to this point, the analysis has been completely general. We now restrict our attention to the case where all ABSs (at a given energy) have momentum $k_x \approx \pm k_F$. This is generally satisfied for energies $\varepsilon$ away from the Andreev levels deep inside the Fermi pocket, as shown in Fig.\ \ref{fig:critical-spectrum}(b). These ABSs are associated to a Fermi surface critical point at $k_x=+k_F$, and given our definition for the critical point (with $\hat{\zeta}=+\hat{x}$), right movers at $k_x \approx +k_F$ are always electronlike while left movers are holelike. Then either $\mathcal T$ or $\mathcal M_x$ imply that the electronlike (holelike) ABSs at  $k_x \approx - k_F$ are left (right) movers. Furthermore, in the \emph{adiabatic} limit where the width of the junction varies slowly with respect to $k_F^{-1}$, scattering between states at $+k_F$ and $-k_F$ is negligible. Note that there are still normal and Andreev reflections in the zero transmission lead subspace, but these do not affect the transmission between leads N1 and N2. Therefore the total $S$ matrix of the reduced system takes on a block-diagonal form in the adiabatic limit,
\begin{equation} \label{eq:s}
    s = s_1 \odot s_2 = \left( \begin{array}{cc|cc}
    r_L^{he} & t_L^{hh} & 0 & 0 \\
    t_R^{ee} & r_R^{eh} & 0 & 0 \\
    \hline
    0 & 0 & r_L^{eh} & t_L^{ee} \\
    0 & 0 & t_R^{hh} & r_R^{he} \\
    \end{array} \right),
\end{equation}
where $\odot$ stands for the composition of the reduced scattering matrices, see Fig.\ \ref{fig:reduced-smatrix}. The upper-left (lower-right) block is designated to the $+k_F$ ($-k_F$) modes, the subscript $L/R$ indicates the outgoing direction, and the superscript $\alpha\beta$ indicates the conversion of a $\beta$-like mode into an $\alpha$-like mode. Here the reflection and transmission matrices in Eq.\ \eqref{eq:s} are related to those in $S_1$ and $S_2$. For example, 
\begin{subequations}
\begin{align}
    \begin{pmatrix} t_R^{ee} & t_R^{eh} \\ t_R^{he} & t_R^{hh} \end{pmatrix} & = t_{2R} \left[ \mathds 1_N - r_{1R} r_{2L} \right]^{-1} t_{1R}, \label{eq:tR} \\
    \begin{pmatrix} r_L^{ee} & r_L^{eh} \\ r_L^{he} & t_L^{hh} \end{pmatrix} & = r_{1L} + t_{1L} r_{2L} \left[ \mathds 1_N - r_{1R} r_{2L} \right]^{-1} t_{1R}. \label{eq:rL}
\end{align}
\end{subequations}
Notice that in Eq.\ \eqref{eq:s} we have assumed that the amplitudes $t_R^{eh}$ and $t_R^{he}$ in Eq.\ \eqref{eq:tR}, as well as $r_L^{ee}$ and $r_L^{hh}$ in Eq.\ \eqref{eq:rL}, all vanish due to the suppressed scattering across the Fermi surface. 

The nonlocal differential conductance at zero temperature is now given by
\begin{subequations}
\begin{align}
    G_{21}(+V_1) & = \left. \frac{dI_2}{dV_1} \right|_{+V_1} = \frac{e^2}{h} \, \mathrm{Tr} \left[ \left( t_R^{ee} \right)^\dag t_R^{ee} \right]_{\varepsilon=eV_1}, \\
    G_{21}(-V_1) & = \left. \frac{dI_2}{dV_1} \right|_{-V_1} = \frac{e^2}{h} \, \mathrm{Tr} \left[ \left( t_R^{hh} \right)^\dag t_R^{hh} \right]_{\varepsilon=eV_1}, \label{eq:smatrix_g}
\end{align}
\end{subequations}
where we used the electron-hole symmetry of the BdG Hamiltonian in the last line. Unitarity of $s$ implies
\begin{subequations}
\begin{align}
    \left( r_L^{he} \right)^\dag r_L^{he} + \left( t_R^{ee} \right)^\dag t_R^{ee} & = \mathds 1_{N_e}, \\
    r_L^{he} \left( r_L^{he} \right)^\dag + t_L^{hh} \left( t_L^{hh} \right)^\dag & = \mathds 1_{N_h},
\end{align}
\end{subequations}
with $N=N_e+N_h$, and $N_e$ ($N_h$) the number of right (left) movers at $+k_F$ (see Fig.\ \ref{fig:reduced-smatrix}). Taking the trace and subtracting yields
\begin{equation}
    \mathrm{Tr} \left[ \left( t_R^{ee} \right)^\dag t_R^{ee} \right] - \mathrm{Tr} \left[ \left( t_L^{hh} \right)^\dag t_L^{hh} \right] = N_e - N_h,
\end{equation}
which are all evaluated at energy $\varepsilon=eV_1$. Finally, when either $\mathcal T$ or $\mathcal M_x$ is preserved, we may interchange $R$ and $L$ in Eq.\ \eqref{eq:smatrix_g}, so that the rectified conductance becomes
\begin{equation} \label{eq:deltaG}
    \delta G_{21}(V_1) = \frac{e^2}{h} \left( N_e - N_h \, \right)_{\varepsilon=eV_1}.
\end{equation}
The validity of this result is independent of the number of ABSs in the junction and depends only on the assumption that scattering across the Fermi surface is strongly suppressed for the nonzero transmission subspace, as well as the presence of $\mathcal T$ or $\mathcal M_x$ \footnote{When both $\mathcal T$ and $\mathcal M_x$ are present, we have $\delta G_{21}(V_1) = -\delta G_{11}(V_1) \equiv G_{11}(-V_1) - G_{11}(V_1)$, so the Fermi sea topology can also be probed via the rectified \emph{local} conductance in that case. See Appendix \ref{app:landauer} for more discussion.}.  We thus see that the rectified conductance $\delta G_{21}$ is still quantized for a wide Andreev junction, provided that the applied bias does not excite any Andreev states deep inside the Fermi pocket. In particular, there always exists a window of small bias $eV_1>0$ such that $N_e - N_h = c_e -c_h = \chi_F$. Upon increasing the bias above the deep Andreev level spacing $v_F/W$, one obtains additional quantized plateaus in $\delta G_{21}$ at larger values.

In the narrow-junction limit ($W \ll |2m^*\Delta_0|^{-1/2}$) the number of positive energy electronlike (holelike) ABSs at $+k_F$ is given by the number of convex (concave) critical points of the Fermi surface $c_e$ ($c_h$) defined relative to the transport direction. As this is independent of the subgap energy we obtain $\delta G_{21} \rightarrow  \sgn(eV_1) (e^2/h) \chi_F$ for a narrow junction in accordance with Eq.\ \eqref{eq:keyresult2}.

\section{Toy model} \label{sec:model}
\begin{figure}
    \centering
    \includegraphics[width=\linewidth]{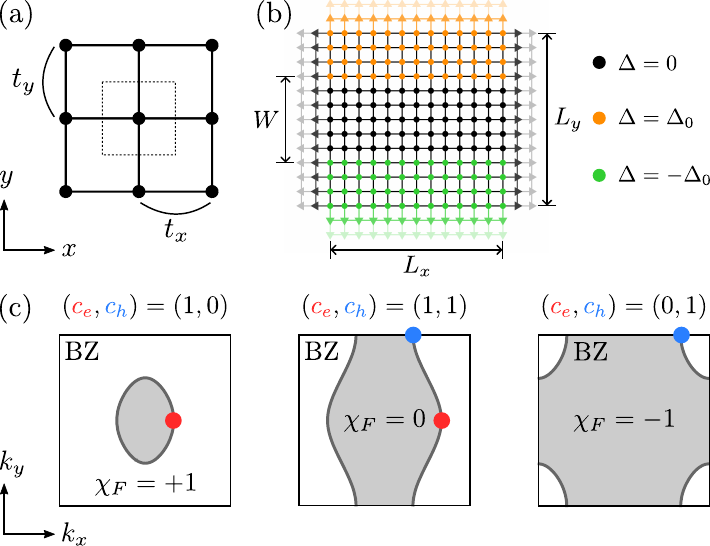}
    \caption{(a) Square lattice with anistropic hopping amplitudes $t_x$ and $t_y$. (b) Setup for the \textsc{Kwant} calculation for a small system ($13 \times 13$ sites). The scattering region with sites shown as dots, whose color indicate the pair potential $\Delta_{\bm n}$, is connected to normal (left and right) and superconducting (bottom and top) leads with sites shown as triangles. (c) Fermi sea in the Brillouin zone (BZ) for $t_y/t_x=0.5$. From left to right: $\mu/t_x = 1$, $\mu/t_x = 3$, and $\mu/t_x = 5$. Here $\chi_F$ and $c_{e,h}$ are labeled for a given spin sector.}
    \label{fig:square}
\end{figure}

To demonstrate and verify our results, we perform transport simulations with the \textsc{Kwant} package \cite{Groth2014,codes}. We first consider a toy model on a square lattice (with lattice constant $a$) with a single $s$ orbital per site, as illustrated in Fig.\ \ref{fig:square}(a). The
corresponding setup for the transport calculation is shown in Fig.\ \ref{fig:square}(b), where a small system is depicted for clarity. The scattering region is connected to two normal leads on the left (N1) and right (N2), and to two proximitized superconducting leads (S) on the top ($\Delta = \Delta_0 > 0 $ ) and bottom ($\Delta = -\Delta_0$).

We consider a model with only nearest-neighbor hopping, with amplitudes $t_x$ and $t_y$ along the $x$ and $y$ directions, respectively. The mean-field Hamiltonian is then given by $\hat H = \hat H_0 + \hat H_{\Delta}$, with
\begin{align}
    \begin{split} \label{eq:toyH0}
        \hat H_0 & = \sum_{\sigma=\uparrow, \downarrow} \sum_{\bm n} \Bigg[ \left( 2t_x + 2t_y -\mu \right) \hat c_{\bm n, \sigma}^\dag \hat c_{\bm n, \sigma} \\
        & - \left( t_x \hat c_{\bm n+\bm e_x, \sigma}^\dag \hat c_{\bm n, \sigma} + t_y \hat c_{\bm n+\bm e_y, \sigma}^\dag \hat c_{\bm n, \sigma} + \mathrm{h.c.} \right) \Bigg], 
    \end{split} \\
    \hat H_{\Delta} & = \sum_{\bm n} \left( \Delta_{\bm n} \hat c_{\bm n, \uparrow}^\dag \hat c_{\bm n, \downarrow}^\dag + \Delta_{\bm n}^* \hat c_{\bm n, \downarrow} \hat c_{\bm n, \uparrow} \right),
\end{align}
where $\hat c_{\bm n,\sigma}^\dag$ ($\hat c_{\bm n,\sigma}$) creates (destroys) an electron with spin $\sigma$ on site $\bm n = (n_x, n_y) \in \mathds{Z}^2$, $\Delta_{\bm n}$ is the superconducting pair potential at site $\bm n$, and $\mu$ is the chemical potential. The normal Hamiltonian $\hat{H}_0$ has dispersion
\begin{equation}
    E(\bm k) = 2t_x \left( 1 - \cos{k_x a} \right) + 2t_y \left( 1 - \cos{k_y a} \right) - \mu, 
\end{equation}
where the Fermi sea is defined by $E(\bm k) \leq 0$. In the following, we consider anisotropic hopping amplitudes with $t_y/t_x=0.5$, and $t_x>0$. In this way, we can access three types of Fermi sea topology:
\begin{equation}
    \chi_F = \begin{cases} +1 & \quad \text{for}\;\; 0 < \mu/t_x < 2, \\
    0 & \quad \text{for}\;\; 2 < \mu/t_x < 4, \\
    -1 & \quad \text{for}\;\;  4 < \mu/t_x < 6. \end{cases}
\end{equation}
Representative cases are illustrated in Fig.\ \ref{fig:square}(c). The count of Fermi surface critical points is $(c_e,c_h) = (1,0)$, $(1,1)$, $(0,1)$, respective to $\chi_F = +1$, $0$, $-1$. Since the normal metal considered here possesses a spin-degenerate Fermi sea, both $\chi_F$ and $c_{e,h}$ in this section are counted per spin.

As explained in the previous section, information on the Fermi sea topology is contained in the nonlocal conductance, which can be computed as (see Appendix \ref{app:landauer})
\begin{align}
    G_{21} = \frac{dI_2}{dV_1} & = \frac{e^2}{h} \int d\varepsilon \left( T_{21}^{ee} - T_{21}^{he} \right) \left( - \frac{df_1}{d\varepsilon} \right) \\
    \overset{T \rightarrow 0}&{=} \frac{e^2}{h} \left( T_{21}^{ee} - T_{21}^{he}  \, \right)_{\varepsilon=eV_1}, 
\end{align}
where $f_1 = f_0(\varepsilon - eV_1)$ is the Fermi distribution in lead N1. The total transmission functions $T_{ij}^{\alpha\beta}(\varepsilon)$ for a charge carrier of type $\beta$ to be transmitted from lead $\text{N}j$ to lead $\text{N}i$ as a charge carrier of type $\alpha$ ($\alpha, \beta = e,h$ and $i,j=1,2$) are calculated in the BdG formalism using \textsc{Kwant}. In the rest of the work, we focus on transport properties at temperature $T=0$. Quantized \textit{plateaus} in $G_{21}(V_1)$ and $\delta G_{21}(V_1)$ for subgap bias $|eV_1|<\Delta_0$ shall remain at finite temperatures provided $k_B T \ll \Delta_0$. 


\begin{figure}[b]
    \centering
    \includegraphics[width=0.75\linewidth]{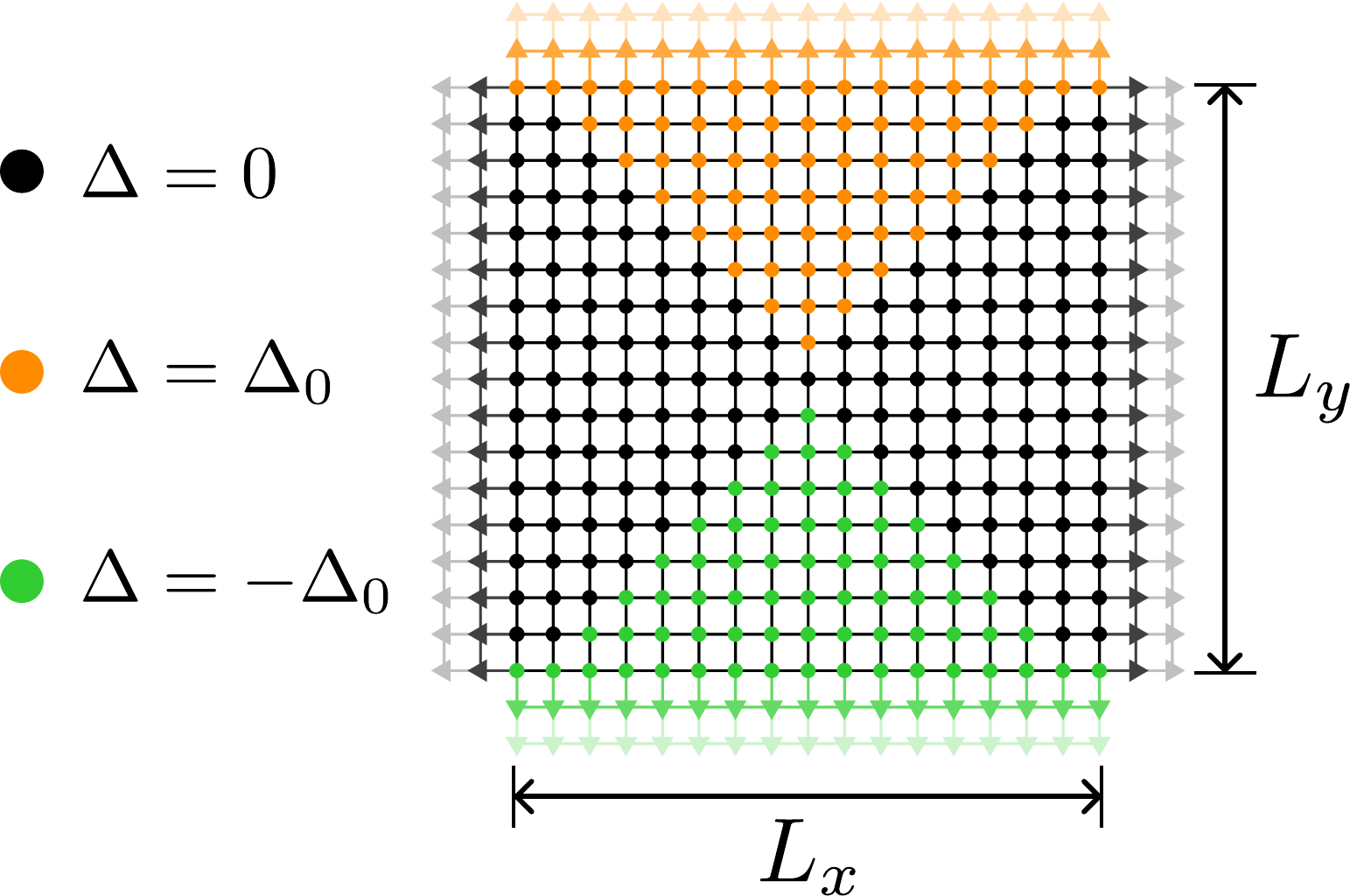}
    \caption{Transport setup for the Andreev point contact. The normal metal is pinched off by two superconducting leads that meet at a \textit{point} (i.e., a region much smaller than $\xi$), while the entire scattering region has dimensions $L_x, L_y \gg \xi$.}
    \label{fig:point-contact}
\end{figure}
 
We now move on to discuss the numerical results for three kinds of junction geometries: (A) narrow and long Andreev junctions; (B) wide and long Andreev junctions; and (C) Andreev point contacts, where the latter are defined in Fig. \ref{fig:point-contact}. Here a long junction has by definition a length $L \gg \xi$, where $\xi$ is the superconducting coherence length. We take $\Delta_0 = 0.1t_x$ and estimate
\begin{equation}
    \xi = \frac{\hbar v_F}{\pi \Delta_0} \sim \frac{t_xa}{\Delta_0} = 10a.
\end{equation}
A narrow junction is defined as hosting a single dispersive ABS per Fermi surface critical point, see Fig.\ \ref{fig:critical-spectrum}(a), while for a wide junction there are multiple ABSs, see Fig.\ \ref{fig:critical-spectrum}(b).

In the following, we take a scattering region of dimensions $L_x \times L_y$ with $L_x=L_y=100a$. For the cases (A) and (B), the linear junction has length $L = L_x$.

\subsection{Narrow Andreev junction}

\begin{figure}[b!]
    \centering
    \includegraphics[width=\linewidth]{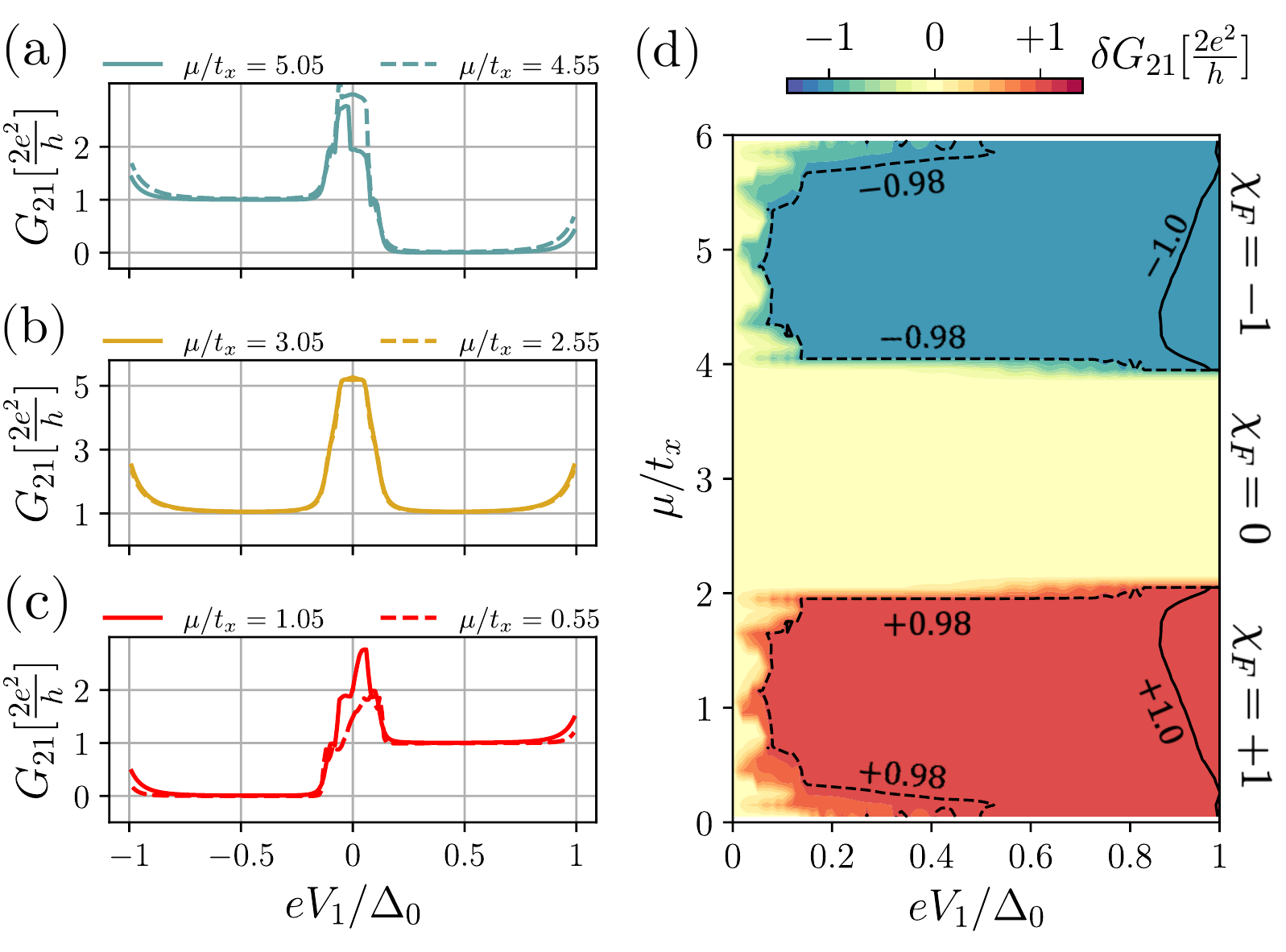}
    \caption{Nonlocal transport in a narrow and long Andreev junction ($W = 6a$, $L_x=L_y=100a$). (a,b,c) Nonlocal conductance $G_{21}$ versus the subgap bias voltage $eV_1$, for three sets of chemical potentials $\mu$ corresponding to $\chi_F = -1$, $0$, and $+1$, respectively. (d) Rectified conductance $\delta G_{21}$ as a function of $\mu$ and $eV_1$. Dashed and solid lines are contours labeled by their respective values. Both $G_{21}$ and $\delta G_{21}$ show robust quantization indicating the Fermi sea geometry and topology, respectively.}
    \label{fig:nj_toymodel}
\end{figure}

\begin{figure*}
    \centering
    \includegraphics[width=0.97\linewidth]{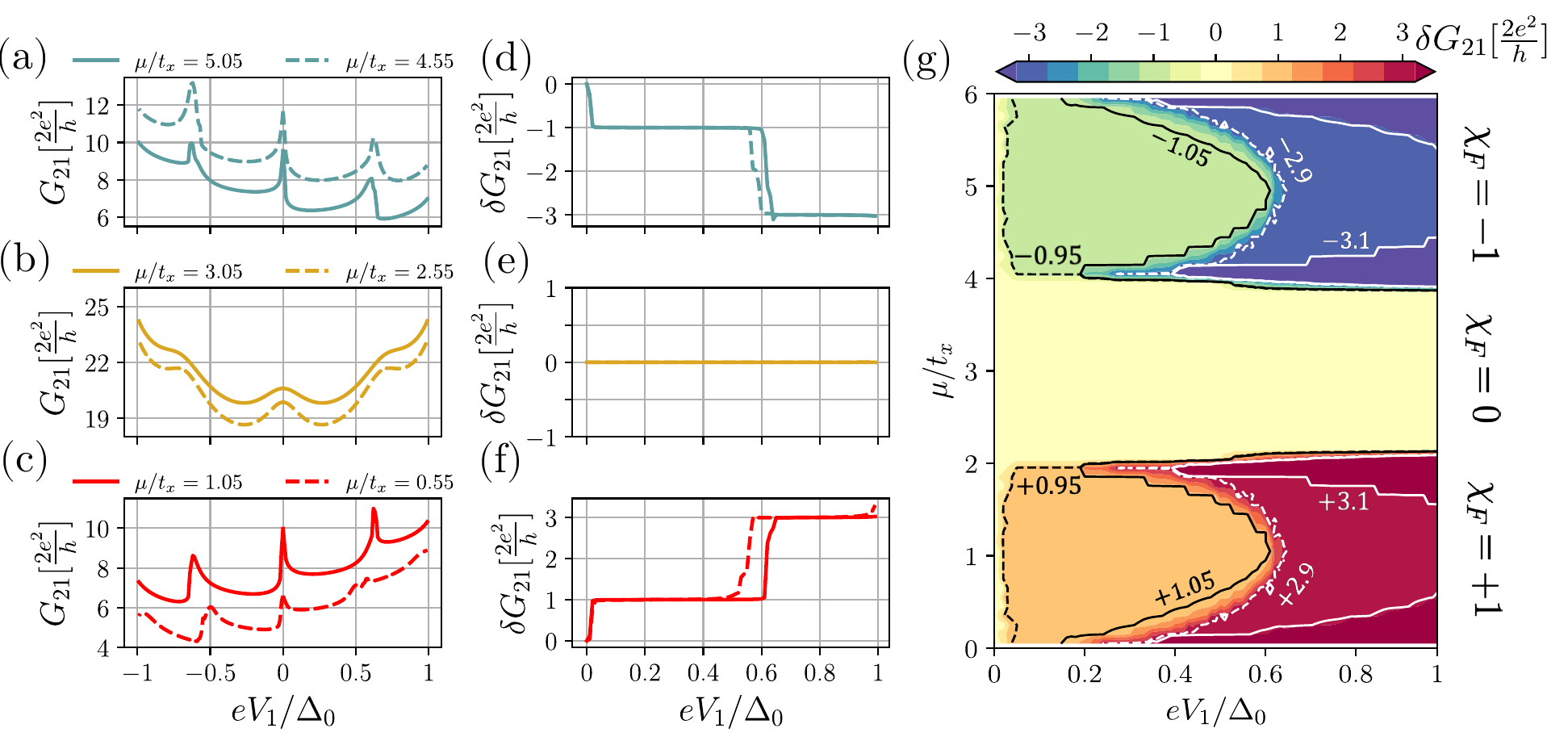}
    \caption{Nonlocal transport in a wide and long Andreev junction ($W = 40a$, $L_x=L_y=100a$). (a,b,c) Nonlocal conductance $G_{21}$ as a function of subgap bias voltage $eV_1$ for three sets of chemical potentials $\mu$ corresponding to $\chi_F=-1$, $0$, and $+1$, respectively. (d,e,f) Rectified conductance $\delta G_{21}$ as a function of $eV_1$ for the same $\mu$ as in (a,b,c), respectively. (g) Color plot of $\delta G_{21}$ as a function of $eV_1$ and $\mu$. Dashed and solid lines are contours labeled by their respective values. While $G_{21}$ is non-universal (i.e., sensitive to $\mu$ for fixed $\chi_F$), $\delta G_{21}$ remains quantized, distinguishing different Fermi sea topologies.}
    \label{fig:wj_toymodel}
\end{figure*}

For a long junction, both crossed Andreev reflection and electron tunneling are suppressed. The only transmission channel from lead N1 to lead N2 is through the ABSs which disperse along the junction. Here we consider a normal region of width $W=6a \lesssim \xi$. As we have demonstrated in Sec.\ \ref{sec:review}, each Fermi surface critical point is associated to \emph{one} pair of ABSs for a narrow Andreev junction, see Fig.\ \ref{fig:critical-spectrum}(a). This is the regime where we expect the theoretical results, given by Eq.\ \eqref{eq:keyresult1} and Eq.\ \eqref{eq:keyresult2}, to be applicable.

The numerical results confirm our predictions. As shown in Figs.\ \ref{fig:nj_toymodel}(a)--(c), the nonlocal conductance $G_{21}$ is quantized (in units of $2e^2/h$) to the integral value $c_e$ (or $c_h$) for subgap bias voltage $eV_1 > 0$ (or $eV_1<0$). The quantized conductance is contributed by the dispersive ABSs close to the Fermi surface critical points (with $k_x \approx \pm k_F$), while the non-universal zero-bias peak originates from the zero-energy ABSs deep inside the Fermi sea (with $k_x \approx 0$). The quantized plateaus at finite bias are universal in the sense that they depend only on the counting of Fermi surface critical points, but not on the detailed shape of the Fermi sea (such as its size). When $\chi_F = c_e -c_h \neq 0$, the nonlocal conductance is \emph{asymmetric} in the bias voltage $V_1$. This asymmetry is quantified by the \emph{rectified conductance} $\delta G_{21}$, as defined in Eq.\ \eqref{eq:def_deltaG}, and is color-plotted as a function of $\mu/t_x$ and $eV_1/\Delta_0$ in Fig.\ \ref{fig:nj_toymodel}(d). In almost the entire subgap regime, $\delta G_{21}$ is found to be quantized to $\chi_F$ (in units of $2e^2/h$). In this case, both $G_{21}$ and $\delta G_{21}$ serve as good markers for the Fermi sea topology. In contrast, the features near zero bias depend on microscopic details, such as the effect of normal reflections at the NS interfaces, which split the nominal zero modes away from zero energy, similar to Fig.\ \ref{fig:critical-spectrum}(a).    This energy splitting oscillates as a function of the chemical potential with an envelope that scales as $\Delta_0^2/\mu$ for $0 < \mu/t_x < 1$, as can be seen close to zero bias in Fig.\ \ref{fig:nj_toymodel}(d). We have verified that the width of the zero-bias peak matches the ABS splitting for an infinitely-long junction with the same $W$ and $\mu$.

\subsection{Wide Andreev junction}

Next, we simulate a wide normal region with $W=40a \gg \xi$. In this case, there are always \emph{multiple} ABSs at each Fermi surface critical point. The BdG spectrum near a convex critical point then resembles Fig.\ \ref{fig:critical-spectrum}(b). As explained in Sec.\ \ref{sec:smatrix}, backscattering among counterpropagating ABSs (which share opposite electron-hole character) is present even in the adiabatic limit as these modes are grouped around $k_x \approx +k_F$ (or $k_x \approx -k_F$). Consequently, we do not expect $G_{21}$ to be quantized. Even when these scattering events are negligible, the number of ABSs available for transmission would depend on microscopic details, such as the value of the chemical potential (for fixed $\chi_F$) and the bias voltage in the lead. Nevertheless, as long as scattering across the Fermi surface is negligible, we have argued in Sec.\ \ref{sec:smatrix} that $\delta G_{21}$ remains quantized.

These predictions are verified by the numerical results, which are summarized in Fig.\ \ref{fig:wj_toymodel}. As shown in (a)--(c), $G_{21}$ is no longer quantized to $c_e$ or $c_h$. Moreover, there is a rather strong dependence on both $\mu$ and $eV_1$. Hence $G_{21}$ \emph{per se} no longer indicates the Fermi sea topology. However, the rectification effect remains robust and quantized, depending only on the topology of the Fermi sea. This is illustrated in Figs.\ \ref{fig:wj_toymodel}(d)--(f), where we show $\delta G_{21}$ corresponding to (a)--(c), respectively. We further show $\delta G_{21}$ as a function of $\mu$ and $eV_1$ in Fig.\ \ref{fig:wj_toymodel}(g). For a given Fermi sea topology, irrespective of the precise value of $\mu$, we find that $\delta G_{21}$ always displays a plateau which attains a quantized value of $\chi_F$ (in units of $2e^2/h$). This confirms that $\delta G_{21}$ remains a good marker for the Fermi sea topology for transport along wide junctions as long as scattering between states at $k_x \approx +k_F$ and states at $k_x \approx -k_F$ is suppressed. In our simulation, the interface between the normal lead and the Andreev junction involves a sharp jump in the pair potential on the lattice scale. However, ordinary backscattering (i.e., leaving electron-hole character unchanged) from $+k$ to $-k$ is mediated by the pair potential as a second-order process and thus suppressed by a factor on the order of $(\Delta_0/t_x)^2$.

\begin{figure}[b!]
    \centering
    \includegraphics[width=\linewidth]{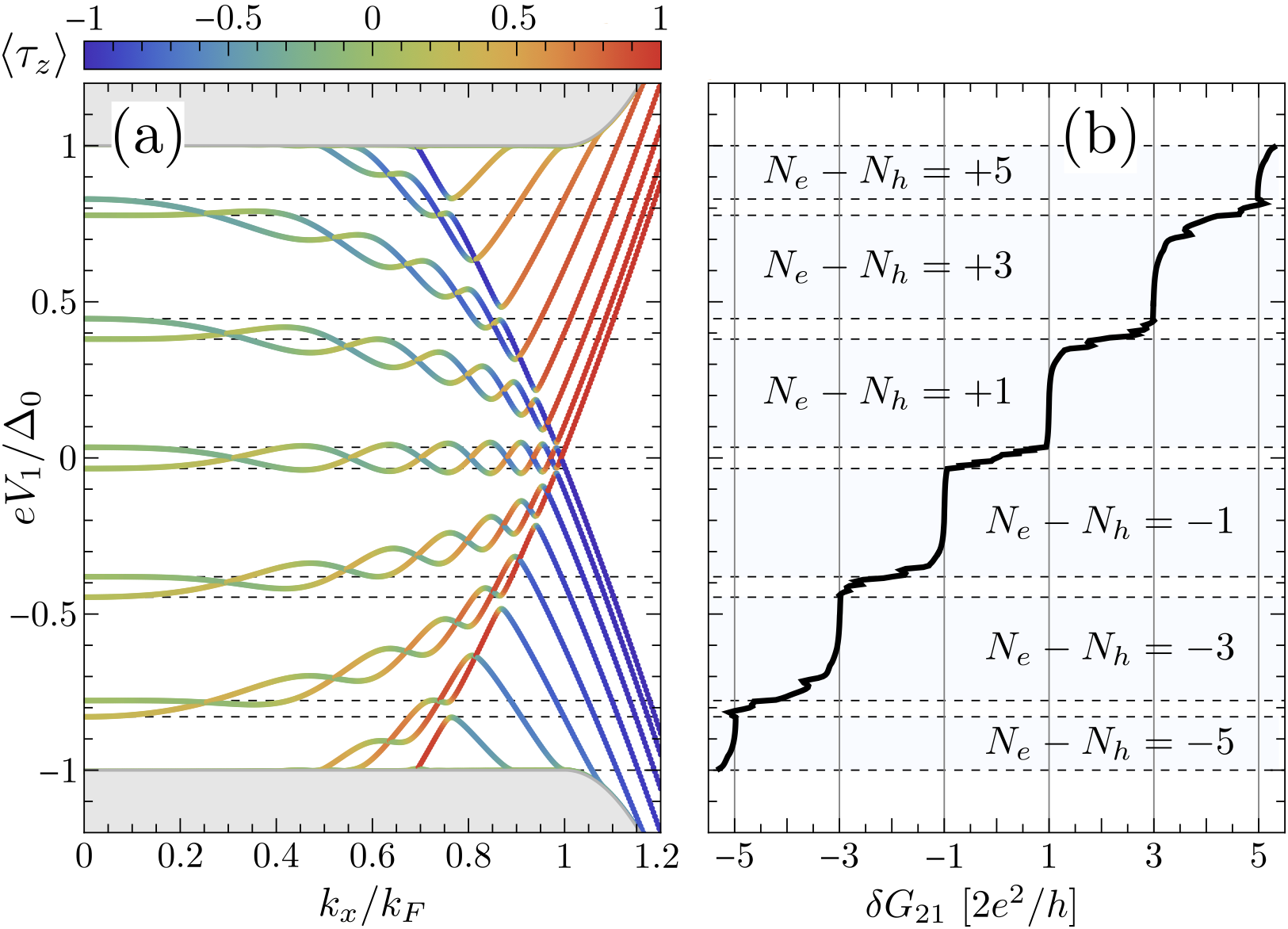}
    \caption{(a) ABS spectrum of an infinitely-long Andreev junction of width $W = 40a$, for $t_y/t_x = 0.5$, $\Delta_0/t_x = 0.1$, and $\mu/t_x = 0.2$. The normal metal has a convex Fermi surface critical point at $\bm k = (k_F,0)$ [see Fig.\ \ref{fig:square}(c)]. Dashed lines indicate the energy of the ABSs at $k_x=0$. (b) Rectified conductance $\delta G_{21}$ as a function of subgap bias voltage $eV_1$, for a scattering region of dimensions $L_x = L_y = 100a$ ($L_x/\xi \approx 36$), and with the same parameters as in (a). Here $N_{e/h}$ is the number of 
    right/left moving 
    modes near the Fermi surface critical point.}
    \label{fig:wide-junction-spectrum}
\end{figure}

The wide Andreev junction hosts additional transport features which are absent in a narrow junction. Particularly, for $\chi_F = \pm 1$, as shown in Fig.\ \ref{fig:wj_toymodel}, there is an extended region in which $\delta G_{21}$ attains a quantized value of $\pm 3$ (in units of $2e^2/h$). This is consistent with our prediction from the $S$-matrix analysis, see Eq.\ \eqref{eq:deltaG}, where $\delta G_{21}(V_1)$ is quantized to an integer that represents the difference between the number of right-moving electronlike ABSs ($N_e$) and the number of left-moving holelike ABSs ($N_h$), at energy $\varepsilon=eV_1$. The plateau transitions happen when $N_e - N_h$ changes its integer value, which correspond to the energies where an extra ABS deep inside the Fermi pocket ($k_x \approx 0$) becomes occupied/depleted. This is  illustrated in Fig.\ \ref{fig:wide-junction-spectrum} for an Andreev junction which hosts five ABSs deep inside the Fermi sea. Correspondingly, the rectified conductance $\delta G_{21}$, shown in Fig.\ \ref{fig:wide-junction-spectrum}(b), exhibits plateaus at $\pm1$, $\pm3$, and $\pm5$ for subgap bias $eV_1 \gtrless 0$, respectively. Note that plateaus appear only at odd integer multiples of $2e^2/h$. This is because the condition for quantization of $\delta G_{21}$ (i.e., the absence of scattering across the Fermi surface) is only satisfied after two extra ABSs are occupied/depleted. If the phase difference across the junction is $0$ instead of $\pi$, so that the topological zero modes are absent, we find similar plateaus, but at even integer multiples of $2e^2/h$.

For $\chi_F = 0$, which corresponds to $2<\mu/t_x<4$, the Fermi sea is surrounded by \emph{open} Fermi surfaces as depicted in the center panel of Fig.\ \ref{fig:square}(c). Hence there are \textit{no} ABSs deep inside the Fermi sea, which implies $N_e=N_h$ for all subgap biases such that no plateau transitions can be found in Fig.\ \ref{fig:wj_toymodel}(e).



\subsection{Andreev point contact}

\begin{figure*}
    \centering
    \includegraphics[width=0.97\linewidth]{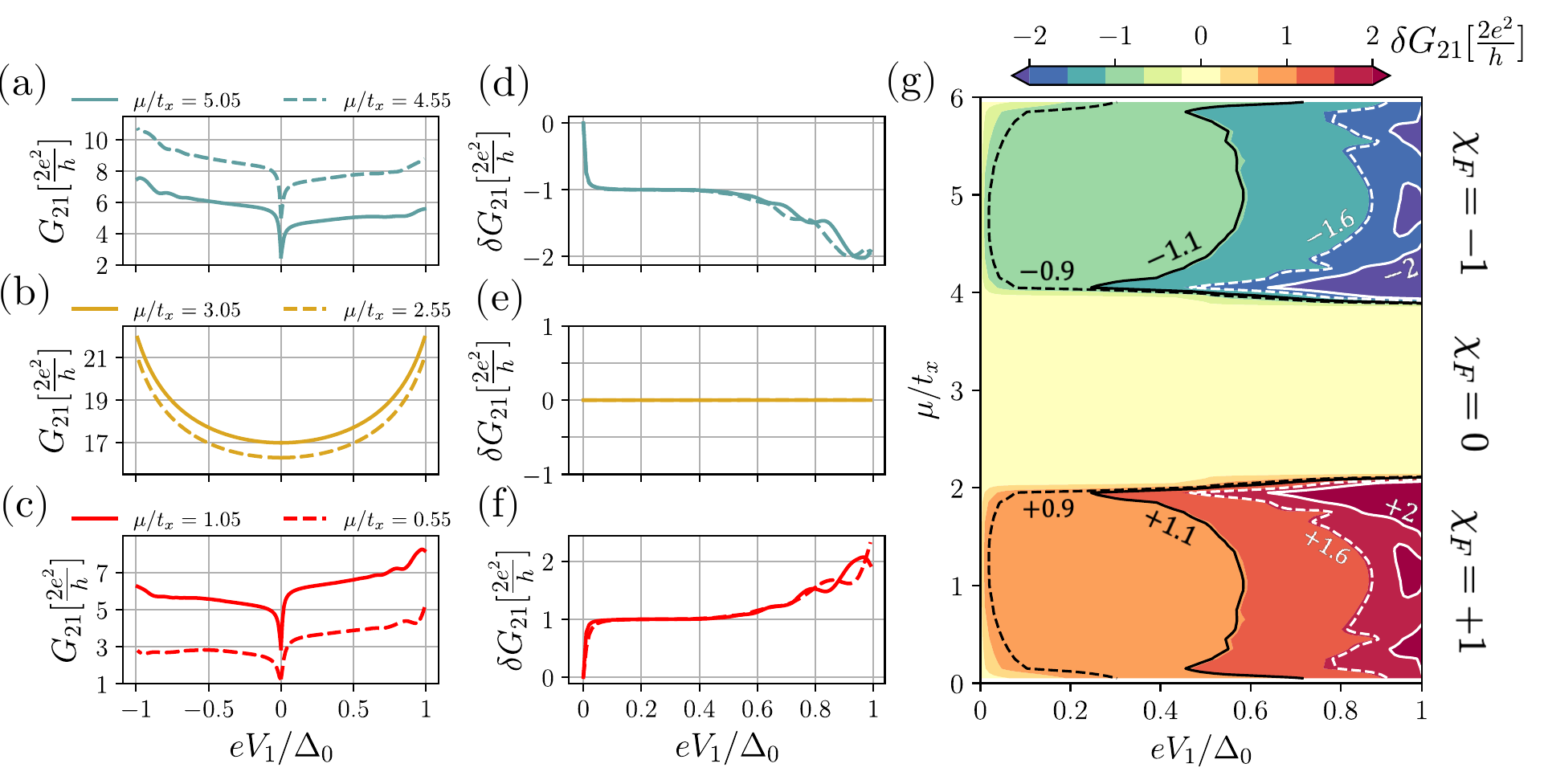}
    \caption{Nonlocal transport in an Andreev point contact ($L_x = L_y = 100a$). (a,b,c) Nonlocal conductance $G_{21}$ as a function of subgap bias voltage $eV_1$ for three sets of chemical potentials $\mu$ corresponding to $\chi_F=-1$, $0$, and $+1$, respectively. (d,e,f) Rectified conductance $\delta G_{21}$ as a function of $eV_1$ for the same $\mu$ as in (a,b,c), respectively. (g) Color plot of $\delta G_{21}$ as a function of $eV_1$ and $\mu$. Dashed and solid lines are contours labeled by their respective values. While $G_{21}$ acquires non-universal behavior, $\delta G_{21}$ remains quantized for small bias voltages, distinguishing different Fermi sea topologies.}
    \label{fig:pc_toymodel}
\end{figure*}

As a final example of our toy model, we consider a point-contact geometry where the normal regions on the two sides of the superconducting constriction are coupled at a point, as depicted in Fig.\ \ref{fig:point-contact}. In contrast to the long Andreev junctions considered before,  neither electron tunneling nor crossed Andreev reflection is suppressed for the Andreev point contact. Hence, we do not expect $G_{21}$ to be universal nor quantized. The numerical results for $G_{21}$, as shown in Fig.\ \ref{fig:pc_toymodel} (a)--(c), indeed exhibit a strong dependence on $\mu$ (for fixed $\chi_F$) and $eV_1$. Nevertheless, an asymmetry with respect to the bias is still present for $\chi_F \neq 0$, and the direction of rectification clearly depends on $\sgn(\chi_F)$.

The theoretical rationale behind the \emph{topological} rectification effect assumes that the ABSs are the only transmission modes across the constriction, and that they \emph{adiabatically} evolve into definite electrons/holes inside the normal leads. These assumptions no longer stand for a point contact. Thus, we do not \emph{a priori} expect any plateau in $\delta G_{21}$ for the Andreev point contact. Remarkably, the numerical results shown in Figs.\ \ref{fig:pc_toymodel}(d)--(f) for $\delta G_{21}$, which correspond to the same parameters used in Figs.\ \ref{fig:pc_toymodel}(a)--(c), as well as the color plot in Fig.\ \ref{fig:pc_toymodel}(g), display a robust quantization of $\delta G_{21}$ in terms of $\chi_F$ for a wide range of parameter values. Hence for a small but finite bias voltage $eV_1$, $\delta G_{21}$ remains a hallmark of the Fermi sea topology for an Andreev point contact.

\section{Materials} \label{sec:materials}

In this section, we discuss numerical results geared towards realistic material platforms. Motivated by recent experimental studies on Josephson junctions \cite{Ren2019, Fornieri2019, Banerjee2022_1, Banerjee2022_2, Banerjee2022_3, Bretheau2017, Wang2018, Park2022}, we focus on a two-dimensional electron gas (2DEG) as realized in InAs quantum wells (with strong spin-orbit coupling) and graphene systems, and examine the feasibility of observing quantized transport in Andreev junctions based on these systems.

\subsection{InAs quantum well}

In recent years, Josephson junctions in hybrid systems based on InAs have been studied extensively with the aim to realize topological superconductivity and Majorana zero modes \cite{Hell2017, Pientka2017,Fornieri2019, Ren2019, Banerjee2022_1}. Moreover, nonlocal conductance measurements have recently been performed on hybrid InAs/Al devices as a probe for topological phase transitions \cite{Banerjee2022_2, Banerjee2022_3}. In the following analysis, we demonstrate that essentially the same setup can be used to probe TAR pertinent to the Fermi sea topology of the Rashba-split 2DEG in an InAs quantum well.

\begin{figure}
    \centering
    \includegraphics[width=\linewidth]{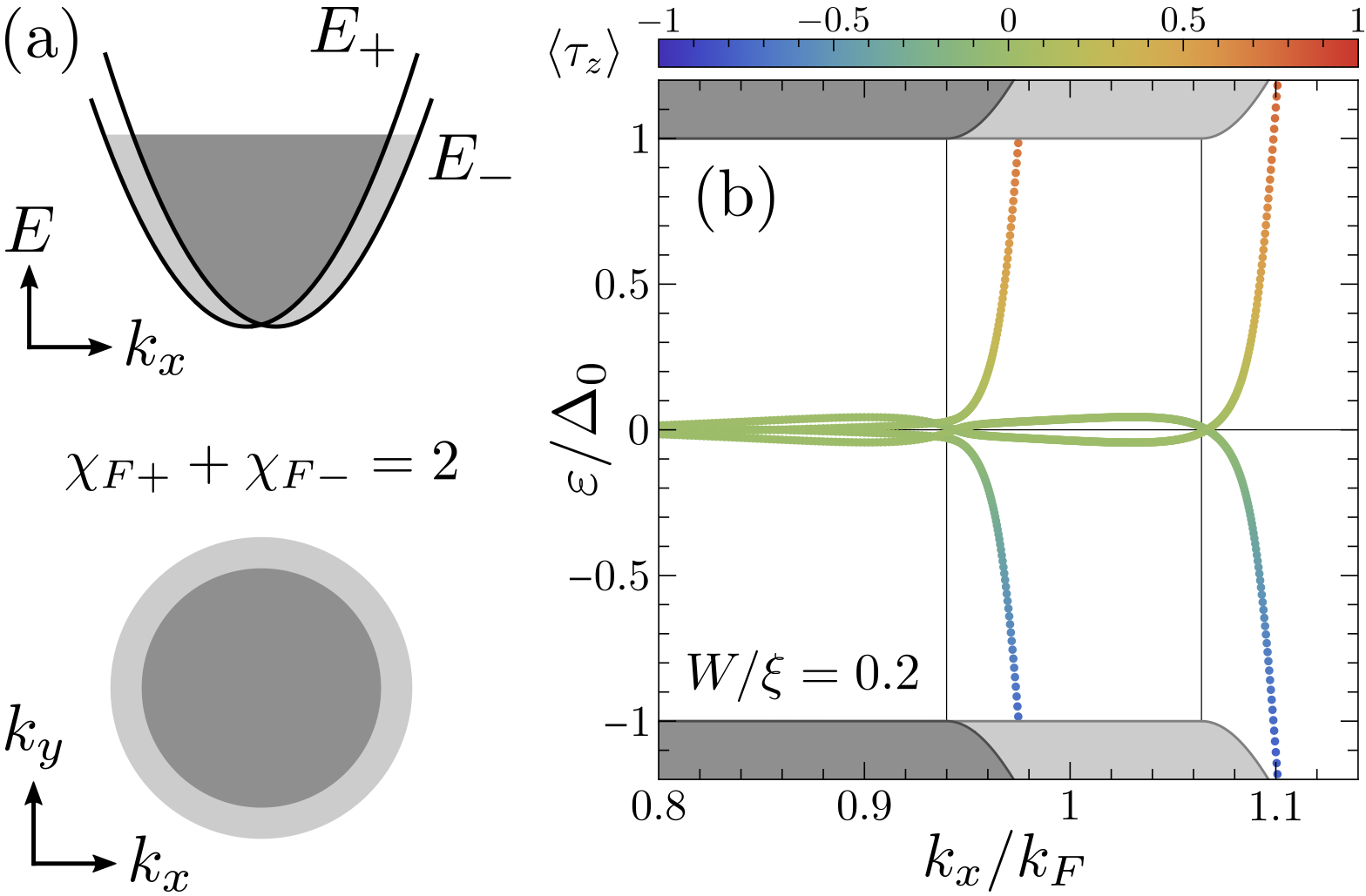}
    \caption{(a) Top: Dispersion close to the band bottom of the Rashba-split 2DEG, with $E_{\pm}(\bm k) = -2t [ \cos(k_x a) + \cos(k_y a) ] \pm 2\lambda_{R}\sqrt{\sin^2(k_x a) + \sin^2(k_y a)}$. Bottom: For our choice of chemical potential, the Fermi sea (shaded in gray) is composed of two electronlike pockets, hence $\chi_F=2$. (b) ABS spectrum of a narrow InAs/Al junction of width $W = 100 \; \text{nm}$ for the parameters in Table \ref{table:InAs_values} and $\mu_N = \mu_S = 10 \; \text{meV}$. Here $k_F$ is the Fermi wave vector for $\lambda_R=0$ and the color indicates the electron-hole character $\left< \tau_z \right>$ of the ABSs.}
    \label{fig:InAs}
\end{figure}

We employ the following tight-binding model on a square lattice (with lattice constant $a= 10 \; \text{nm}$) to simulate an Andreev junction in an InAs/Al system: $\hat{H} = \hat{H}_{0} + \hat{H}_{\Delta}$ with
\begin{align}
    \begin{split} \label{eq:InAsH0}
        & \hat H_0 = \sum_{\sigma=\uparrow, \downarrow} \sum_{\bm n} \left( 4t -\mu_{\bm n} \right) \hat c_{\bm n, \sigma}^\dag \hat c_{\bm n, \sigma} \\
        & - \sum_{\sigma, \sigma'} \sum_{\bm n} \left[ (t\delta_{\sigma,\sigma'}+i\lambda_{R} \sigma^y_{\sigma,\sigma'})\hat c_{\bm n+\bm e_x, \sigma}^\dag \hat c_{\bm n, \sigma'} + \mathrm{h.c.} \right] \\
        & - \sum_{\sigma, \sigma'} \sum_{\bm n} \left[ (t\delta_{\sigma,\sigma'}-i\lambda_{R} \sigma^x_{\sigma,\sigma'})\hat c_{\bm n+{\bm e}_y, \sigma}^\dag \hat c_{\bm n, \sigma'} + \mathrm{h.c.} \right],
    \end{split} \\
    & \hat H_{\Delta} = \sum_{\bm n} \left( \Delta_{\bm n} \hat c_{\bm n, \uparrow}^\dag \hat c_{\bm n, \downarrow}^\dag + \Delta_{\bm n}^* \hat c_{\bm n, \downarrow} \hat c_{\bm n, \uparrow} \right),
\end{align}
where $\hat c_{\bm n,\sigma}^\dag$ ($\hat c_{\bm n,\sigma}$) creates (destroys) an electron with spin $\sigma$ on site $\bm n = (n_x, n_y) \in \mathds{Z}^2$, $\mu_{\bm n}$ and $\Delta_{\bm n}$ are respectively the chemical potential and pair potential on site $\bm n$. The effective hopping amplitude is $t= \hbar^2/(2m^* a^2) = 15\; \text{meV}$, where an effective mass $m^* = 0.025 m_e$ ($m_e$ being the bare electron mass) has been adopted. The Rashba spin-orbit coupling strength is chosen to be $\lambda_R = 0.75 \; \text{meV}$. In the superconducting region proximitized by Al leads [cf.\ Fig.\ \ref{fig:square}(b) and Fig.\ \ref{fig:point-contact}], the magnitude of the pairing gap is chosen to be $\Delta_0=0.15 \; \text{meV}$. These are typical parameter values adopted from recent experimental studies \cite{Banerjee2022_1,Banerjee2022_2, Banerjee2022_3}.

\begin{table}[t!]
\centering
\caption{Parameter values (in meV) used for simulating an Andreev junction in a hybrid InAs/Al system.}
\begin{center}
\begin{tabular}{c c c c c}
\Xhline{1pt}
$t$ & $\lambda_{R}$ & $\Delta_0$ & $\mu_N$ & $\mu_S$\\
\hline
$15.0$ & $0.75$ & $0.15$ & $10.0$ & $10.0$, $10.3$ \\
\Xhline{1pt}
\end{tabular}
\label{table:InAs_values}
\end{center}
\end{table}

\begin{figure}[b]
    \centering
    \includegraphics[width=\linewidth]{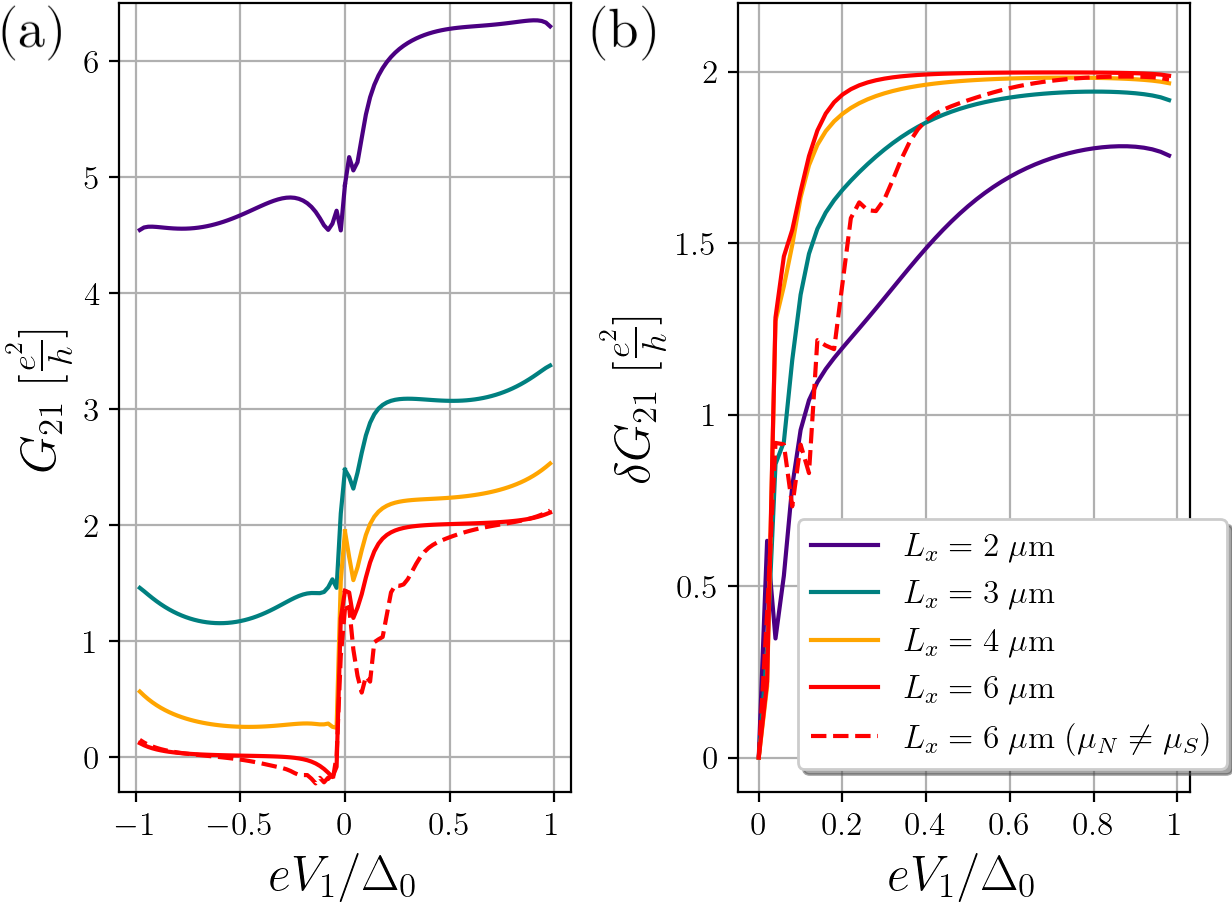}
    \caption{Nonlocal transport in an InAs/Al narrow Andreev junction of various junction lengths $L_x$. (a) Nonlocal conductance $G_{21}$ versus subgap bias $eV_1$ showing plateaus at $2e^2/h$ (for $eV_1 >0$) and $0$ (for $eV_1 <0$) for sufficiently long junctions. (b) Rectified conductance $\delta G_{21}$. Quantized plateaus reflecting $\chi_F = 2$ are consistently observed over a wide range of junction lengths. In the presence of a mismatch in $\mu$ across the NS interface, the topological Andreev rectification effect survives for large enough bias.}
    \label{fig:InAs_nj}
\end{figure}

\begin{figure}
    \centering
    \includegraphics[width=\linewidth]{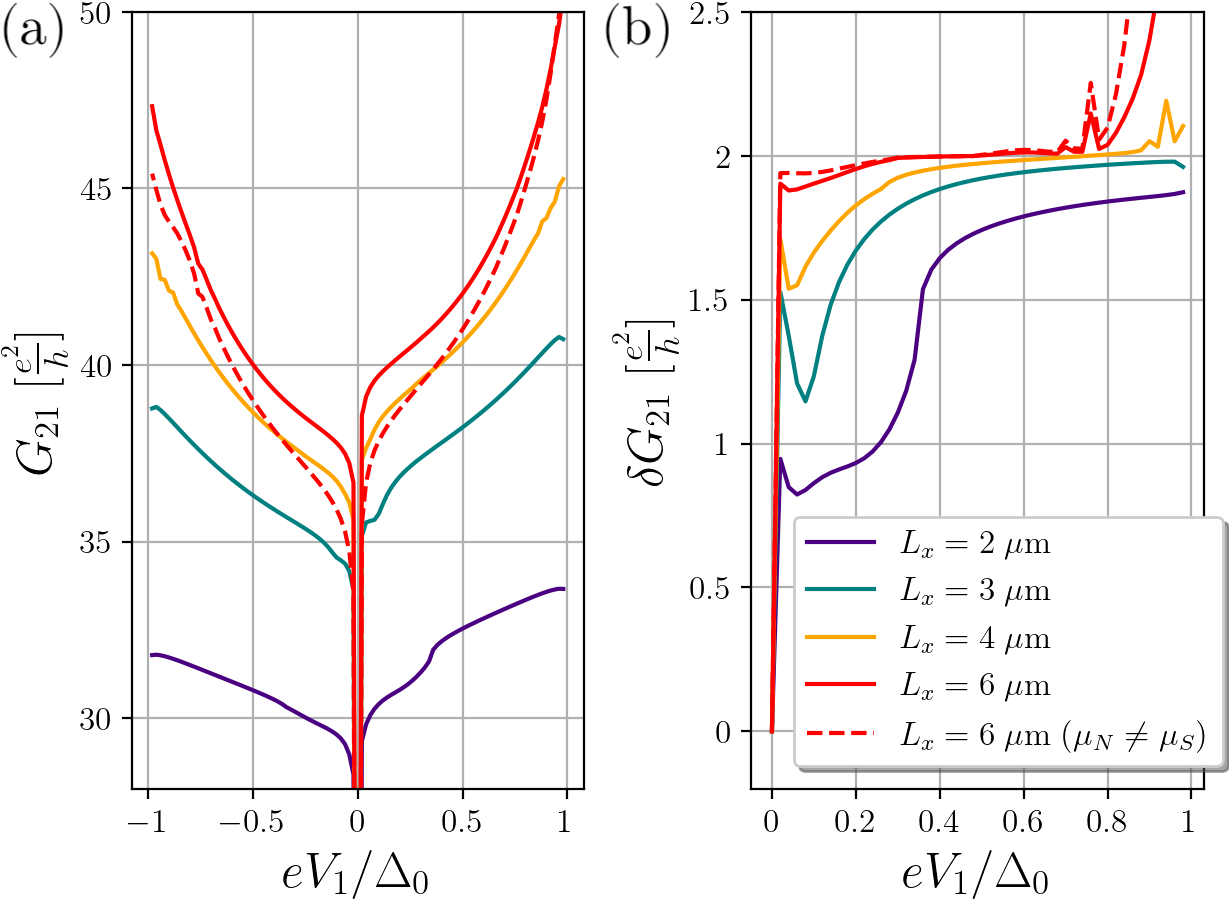}
    \caption{Nonlocal transport in an InAs/Al Andreev point contact, for several values of $L_x=L_y$ that characterizes the size of the scattering region. (a) Nonlocal conductance $G_{21}$ versus subgap bias $eV_1$ showing bias-asymmetry for a wide range of geometries. (b) Rectified conductance $\delta G_{21}$, which captures the bias-asymmetry, exhibits quantized plateaus that reflect the Fermi sea topology ($\chi_F = 2$) for large enough bias and $L_x \gg \xi$. Topological rectification is stable against mismatch in $\mu$ across the NS interface.}
    \label{fig:InAs_pc}
\end{figure}

The chemical potential in the normal region (including normal leads) is fixed at $\mu_N= 10 \; \text{meV}$, which gives rise to two electronlike Fermi seas as depicted in Fig.\ \ref{fig:InAs}(a). In practice, the chemical potential can be tuned by gating within a range of $\sim 1-100 \; \text{meV}$. For most calculations presented below, we set the chemical potential in the superconducting regions $\mu_S$ equal to $\mu_N$. Given $\mu \sim 10 \; \text{meV}$, the superconducting coherence length $\xi \sim 500 \; \text{nm}$, which matches the experimental values reported in Refs.\ \cite{Banerjee2022_1,Banerjee2022_2, Banerjee2022_3}. The parameter values adopted in our \textsc{Kwant} simulation are summarized in Table \ref{table:InAs_values}.

We first simulate a narrow Andreev junction where the normal region has width $W= 100 \; \text{nm}$. This is a realistic value based on recent experiments \cite{Banerjee2022_1,Banerjee2022_2,Banerjee2022_3}. The setup follows Fig.\ \ref{fig:square}(b). The junction is attached to normal leads of width $L_y = 1 \; \mathrm{\mu m}$, and we consider junctions with lengths varying from $2 \; \mathrm{\mu m}$ to $6 \; \mathrm{\mu m}$. The corresponding numerical results are summarized in Fig.\ \ref{fig:InAs_nj}. While sufficiently long junctions ($L_x \gtrsim 6 \; \mathrm{\mu m}$) are required for observing quantized plateaus in $G_{21}$ that reflect the geometry of the Fermi sea $(c_e=2, \;  c_h=0)$, a shorter junction ($L_x \gtrsim 3 \; \mathrm{\mu m}$) already exhibits quantization in the rectified conductance $\delta G_{21}$ which encodes the topology of the Fermi sea ($\chi_F=2$).

In a real device, the superconducting region is most likely at a different chemical potential than the normal region, as the Al contacts locally dope the 2DEG. Such an electrostatic gradient decreases the transparency of the junction by enhancing ordinary reflections. This results in an increased energy splitting of the nominal Andreev zero modes. Thus, a quantized response is anticipated only for biases $|eV_1|$ above this energy scale. To model this effect, we consider a mismatch in the chemical potential across the NS interface that is sharp on the lattice scale (with $\mu_N = 10 \; \mathrm{meV}$ and $\mu_S = 10.3 \; \mathrm{meV}$). Our simulations shown in Fig.\ \ref{fig:InAs_nj} demonstrate that the rectification and the concomitant signatures of the Fermi sea topology in the nonlocal conductance are stable for a step mismatch $\mu_S - \mu_N \sim \Delta_0$. While in practice the mismatch may be larger than $\Delta_0$, a value of the order of $\Delta_0$ for a \emph{step mismatch} serves as a proxy for a more realistic potential profile that varies slowly on the scale of the Fermi wavelength $\lambda_F$. In this case, the transparency of the NS interface is only slightly reduced and the TAR effect is expected to remain.

Next, we consider an Andreev point contact using the same setup as shown in Fig.\ \ref{fig:point-contact}. Our numerical results are shown in Fig.\ \ref{fig:InAs_pc}. While there is no quantization in the nonlocal conductance $G_{21}$ for the point contact, as expected, the rectification effect remains observable. Moreover, for systems that are large compared to $\xi$, the rectified nonlocal conductance $\delta G_{21}$ is quantized and reflects the Fermi sea topology. This conclusion is unchanged when there is a mismatch in the chemical potential across the NS interface that is sharp on the lattice scale and of the order of $\Delta_0$.

We note that all our numerical results were obtained for disorder-free systems. For the state-of-the-art devices reported in Refs.\ \cite{Banerjee2022_1,Banerjee2022_2, Banerjee2022_3}, an electron mean free path in the range of $\sim 200 - 600 \; \text{nm}$ was realized (i.e., comparable to the superconducting coherence length). We anticipate that with improved fabrication techniques, the mean free path can be further increased to realize a \emph{long} and \emph{ballistic} Andreev junction, in which both quantized $\delta G_{21}$ and $G_{21}$ can be measured.  

\subsection{Graphene}

\begin{figure}[b]
    \centering
    \includegraphics[width=\linewidth]{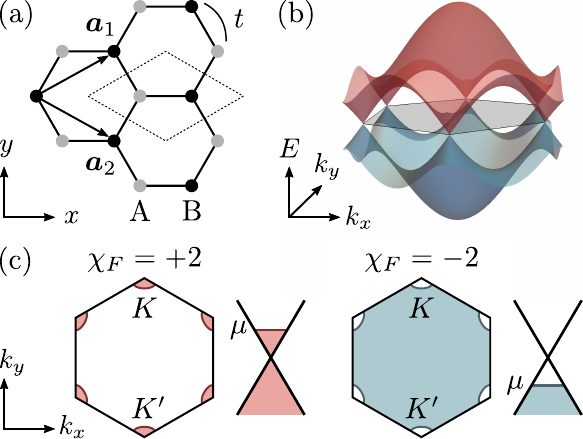}
    \caption{(a) Graphene lattice where the unit cell is the dashed rhombus. (b) Energy bands obtained with the nearest-neighbor model 
    showing the Dirac cones at the BZ corners $K$ and $K'$. (c) Fermi sea for $\mu = \pm 0.5t$  with $\chi_F = \pm 2$ (per spin), together with a sketch of the occupation of the Dirac cones.}
    \label{fig:mlg}
\end{figure}

Another interesting platform for realizing our proposal is graphene. Josephson junctions in graphene have been extensively studied both theoretically and experimentally \cite{Beenakker2006,Titov2006,Heersche2007,Du2008,Calado2015,BenShalom2016,Efetov2016,Allen2016,Amet2016,Bretheau2017,Nanda2017, Wang2018, Li2018,Park2022}. As such, it seems reasonable that nonlocal conductance measurements---to demonstrate the TAR effect and probe the Fermi sea topology---should be achievable in graphene systems.

We first consider monolayer graphene (MLG) and simulate transport in Andreev junctions in the ballistic regime for reasonable device geometries using a scaled honeycomb lattice, and taking realistic values for the proximitized superconducting gap $\Delta_0$. Secondly, we consider Bernal bilayer graphene (BLG) with an interlayer bias. This system has a highly-tunable Fermi sea as a function of the chemical potential and interlayer bias, with $|\chi_F|$ ranging from 0 to 3 (per spin and valley). It is therefore an interesting candidate for investigating TAR.

In all our graphene simulations, we take a scattering region of dimensions $L_x \times L_y$ with $L_x=L_y$, and consider a linear junction of length $L=L_x$ and width $W$, similar to what is shown in Fig. \ref{fig:square}(b) for the square lattice.
Normal (superconducting) leads are attached to the left and right (top and bottom) of the scattering region. 

\subsubsection{Monolayer graphene}

To demonstrate TAR in monolayer graphene (MLG) we consider the simplest lattice model for the charge carriers close to charge neutrality \cite{CastroNeto2009}. To this end, we consider a lattice model $\hat H = \hat H_0 + \hat H_{\Delta}$ with
\begin{widetext}
\begin{align}
    \hat H_0 & = \sum_{\bm R,\sigma} \left[ -\mu \left( \hat a_{\bm R,\sigma}^\dag \hat a_{\bm R,\sigma} + \hat b_{\bm R,\sigma}^\dag \hat b_{\bm R,\sigma} \right) - t \left( \hat a_{\bm R,\sigma}^\dag \hat b_{\bm R,\sigma} + \hat a_{\bm R+\bm a_1,\sigma}^\dag \hat b_{\bm R,\sigma} + \hat a_{\bm R+\bm a_2,\sigma}^\dag \hat b_{\bm R,\sigma} + \text{h.c.} \right) \right], \label{eq:mlgH0} \\
    \hat H_\Delta & = \sum_{\bm R} \left( \Delta_{\bm R} \hat a_{\bm R,\uparrow}^\dag \hat a_{\bm R,\downarrow}^\dag + \Delta_{\bm R+\bm \tau} \hat b_{\bm R,\uparrow}^\dag \hat b_{\bm R,\downarrow}^\dag + \text{h.c.} \right),
\end{align}
\end{widetext}
where $\bm R = n_1\bm a_1 + n_2\bm a_2$ are lattice vectors with $n_{1/2} \in \mathds Z$ and $\bm a_{1/2} = a ( \sqrt{3}/2, \pm 1/2 )$ with $a \approx 0.25 \; \text{nm}$, see Fig.\ \ref{fig:mlg}(a). Here $\hat a_{\bm R,\sigma}^\dag$ and $\hat b_{\bm R,\sigma}^\dag$ ($\hat a_{\bm R,\sigma}$ and $\hat b_{\bm R,\sigma}$) create (destroy) an electron with spin $\sigma = \uparrow,\downarrow$ on sublattice A and B, respectively, at sites $\bm R$ and $\bm R+\bm \tau$ with $\bm \tau = a \hat x / \sqrt{3}$. The nearest-neighbor hopping amplitude is given by $t \approx 2.8$~eV and $\Delta_{\bm r}$ gives the proximity-induced pair potential at position $\bm r$ with $\Delta_{\bm r} = \Delta_0 \left[ \theta(y - W/2) - \theta(-y - W/2) \right]$.

In the low-density regime, the energy bands are given by a pair of spin-degenerate Dirac cones at the two distinct zone corners (valleys) $K$ and $K'$, as shown in Fig.\ \ref{fig:mlg}(b). In this case, the Fermi sea topology (per spin and valley) is given by $\chi_F = \sgn(\mu/t)$ where the count of Fermi surface critical points, per spin and valley, is $(c_e , c_h) = (1 ,0)$ for electron doping and $(c_e, c_h) = (0, 1)$ for hole doping. This is illustrated in Fig.\ \ref{fig:mlg}(c).
 
Using $\Delta_0 = 0.2 \; \text{meV}$ as the proximity-induced gap for Al contacts, we find $\xi / a = \sqrt{3} t / 2 \pi \Delta_0 \sim 4 \times 10^3$. To make the computation feasible for an intermediately long ($L_x \gtrsim \xi$) and narrow ($W \lesssim \xi$) Andreev junction, and since we are only considering the regime with a linear dispersion, we consider a \emph{scaled} honeycomb lattice \cite{Liu2015}. This is equivalent to solving the Dirac equation on a hexagonal grid scaled by a factor $s$, where
\begin{equation}
    a \mapsto s a, \qquad t \mapsto t /s,
\end{equation}
such that $\hbar v_F = \sqrt{3} \, ta/2$ is invariant. This is justified as long as $s \ll |t/\mu|$, or equivalently $\lambda_F \gg sa$, such that we remain in the linear regime. Since we want to simulate a large system (comparable to $\xi \sim 1 \; \mathrm{\mu m}$) we need a sufficiently large scaling factor $s$ which puts an upper bound on $|\mu|$. To this end, we consider carrier densities $|n| = k_F^2/\pi < 10^{11}$~cm$^{-2}$ which, assuming a linear dispersion, yields $|\mu| < 30$~meV and thus $s \ll 100$. In practice, it is more favorable to consider a higher carrier density with $|\mu| \gtrsim 100 \;\text{meV}$, as a spatial variation of the chemical potential $\delta \mu \sim 5 \; \text{meV}$  is present even for high-quality graphene devices on hBN substrates \cite{Xue2011}. When $\mu$ is too small, the presence of electron-hole puddles across the device, which exhibit different Fermi sea topology, can preclude low-energy Andreev states even in a $\pi$ junction \cite{Bretheau2017}. 

In the following, we take $s=12$, which gives a scaled lattice constant $sa$ of about $3$~nm and $\xi/sa \sim 300$. We then numerically implement an Andreev junction of length $L_x = 1000\:sa \approx 3 \;\mathrm{\mu m}$ and width $W = 100\:sa \approx 300$~nm in \textsc{Kwant} \cite{Groth2014,codes}. These dimensions are comparable to those of state-of-the-art devices \cite{BenShalom2016,Bretheau2017,Nanda2017,Wang2018, Park2022}. An overview of the parameters used for our \textsc{Kwant} simulation is given in Table \ref{table:graphene_values}.

\begin{table}
\centering
\caption{Parameter values for simulating an Andreev junction in hybrid graphene--superconductor systems.}
\begin{center}
\begin{tabular}{l c c c c}
\Xhline{1pt}
& $t$ & $t_\perp$ & $\gamma_3$ & $\Delta_0$ \\
\hline
MLG & $2.8 \; \text{eV}$ & n.a. & n.a. & $0.2 \; \text{meV}$ (Al)  \\
& & & & $1 \; \text{meV}$ (Nb or MoRe) \\
\hline 
BLG & $2.8 \; \text{eV}$ & $0.1t$ & $0.1t$ & $0.0025t$ \\
\Xhline{1pt}
\end{tabular}
\label{table:graphene_values}
\end{center}
\end{table}

\begin{figure}
    \centering
    \includegraphics[width=\linewidth]{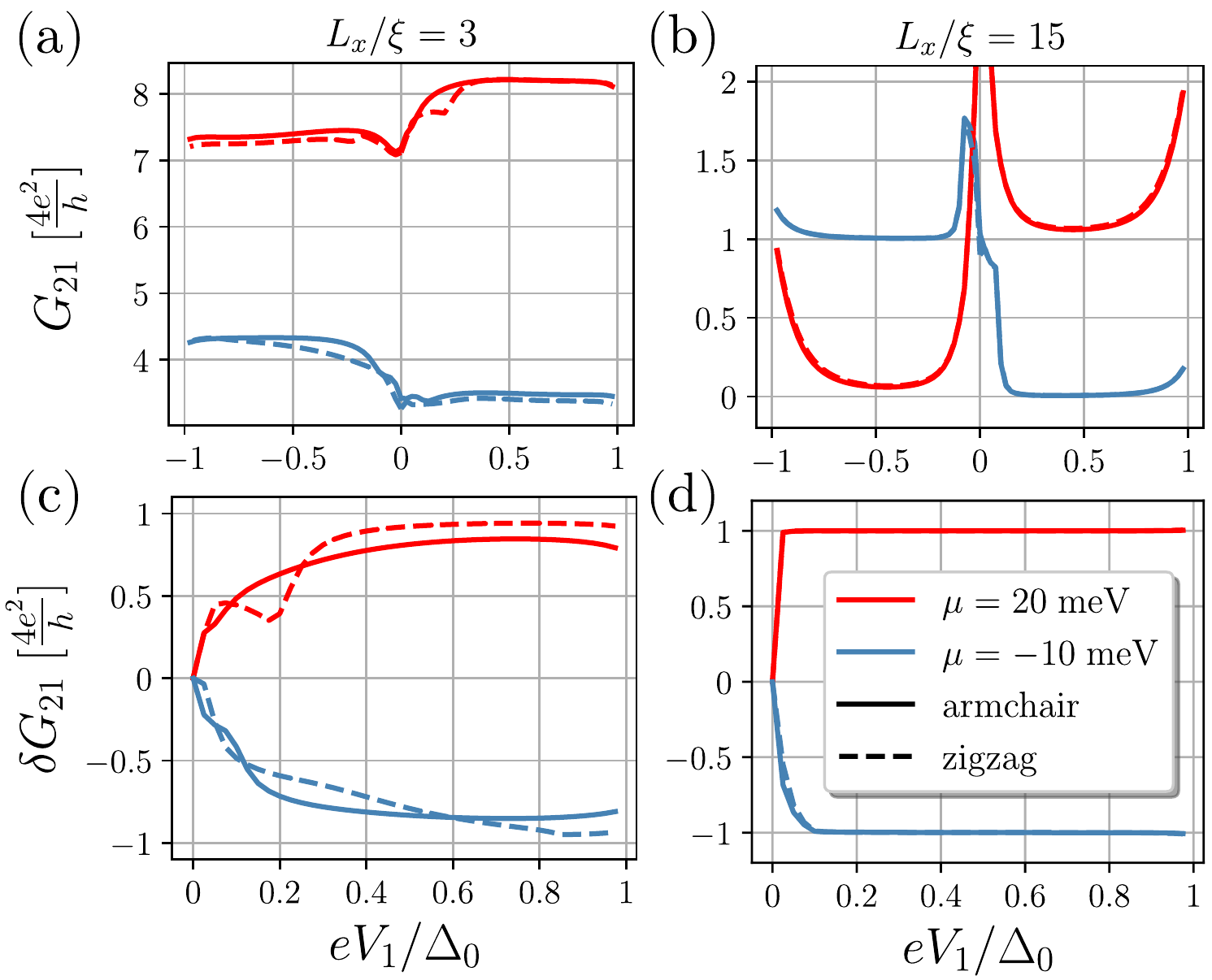}
    \caption{Nonlocal transport in a narrow MLG-based Andreev junction, with $W=300 \; \text{nm}$ and $L_x = 3 \; \mathrm{\mu m}$, for a system of dimensions $L_x \times L_x$ with a scaled lattice constant $sa= 3$~nm. (a) and (b): Nonlocal conductance $G_{21}$ versus the subgap bias $eV_1$ for transport along the armchair (solid) and zigzag (dashed) direction, with $\Delta_0 = 0.2 \; \text{meV}$ (Al) and $\Delta_0 = 1 \; \text{meV}$ (Nb or MoRe), respectively. In the latter case, the system is in the long-junction regime where the quantization of $G_{21}$ reflects the Fermi sea geometry. (c) and (d): Rectified conductance $\delta G_{21}$ as a function of $eV_1$ corresponding to (a) and (b), respectively. In both cases, $\delta G_{21}$ shows quantization indicating the Fermi sea topology.}
    \label{fig:mlg-g}
\end{figure}

The nonlocal conductance $G_{21}$ and the rectified conductance $\delta G_{21}$ are shown in Fig.\ \ref{fig:mlg-g} for two values of the chemical potential corresponding to $\chi_F=\pm 1$ (per spin and valley), and with $\Delta_0 = 0.2 \; \text{meV}$ (Al contacts) and $\Delta_0 = 1 \; \text{meV}$ (Nb or MoRe contacts). We first notice that the qualitative features of the nonlocal response are \emph{insensitive} to whether transport is along the armchair or zigzag direction. This is anticipated, as the Fermi sea has full rotational symmetry in the Dirac regime. For Al contacts [Fig.\ \ref{fig:mlg-g}(a) and Fig.\ \ref{fig:mlg-g}(c)] we have $L_x/\xi = 3$ and the nonlocal conductance $G_{21}$ is \emph{not} quantized to the count of Fermi surface critical points. This can be attributed to the presence of electron-tunneling across the rather short Andreev junction. Nevertheless, $\delta G_{21}$ is found to be nearly quantized to $\pm 1$ (in units of $4e^2/h$) allowing TAR to be observed in this setting. We also considered Nb or MoRe contacts [Fig.\ \ref{fig:mlg-g}(b) and Fig.\ \ref{fig:mlg-g}(d)] with $L_x/\xi = 15$, which is in the long-junction regime. In this case, we find almost perfect quantization in $G_{21}$ which captures the Fermi sea geometry in accordance to Eq.\ \eqref{eq:keyresult0}. For realistic devices with short mean free paths compared to $L_x$ and $\xi$, the quantization in $G_{21}$ and $\delta G_{21}$ would both be degraded, but when the chemical potential is tuned through charge neutrality one still expects to observe a sudden jump in $\delta G_{21}$ which reflects the change in the Fermi sea topology. We thus believe that it is possible to observe signatures of TAR in current state-of-the-art hybrid superconductor-graphene devices.

\subsubsection{Bernal bilayer graphene}

Bernal-stacked bilayer graphene (BLG) is a potentially more interesting platform for probing Fermi sea topology using TAR. Upon the application of an interlayer bias, by varying top and bottom gate voltages, this system hosts a much richer landscape of different Fermi sea topologies as compared to MLG. One can thus potentially tune experimentally between different values of $\chi_F$ by varying the interlayer bias as well as the chemical potential. Moreover, SNS junctions have been realized experimentally in BLG proximitized with Al \cite{Allen2016} ($\Delta_0 \approx 0.2 \; \text{meV}$) as well as NbSe$_2$ thin films \cite{Efetov2016} ($\Delta_0 \approx 1 \; \text{meV}$).

\begin{figure}[t]
    \centering
    \includegraphics[width=\linewidth]{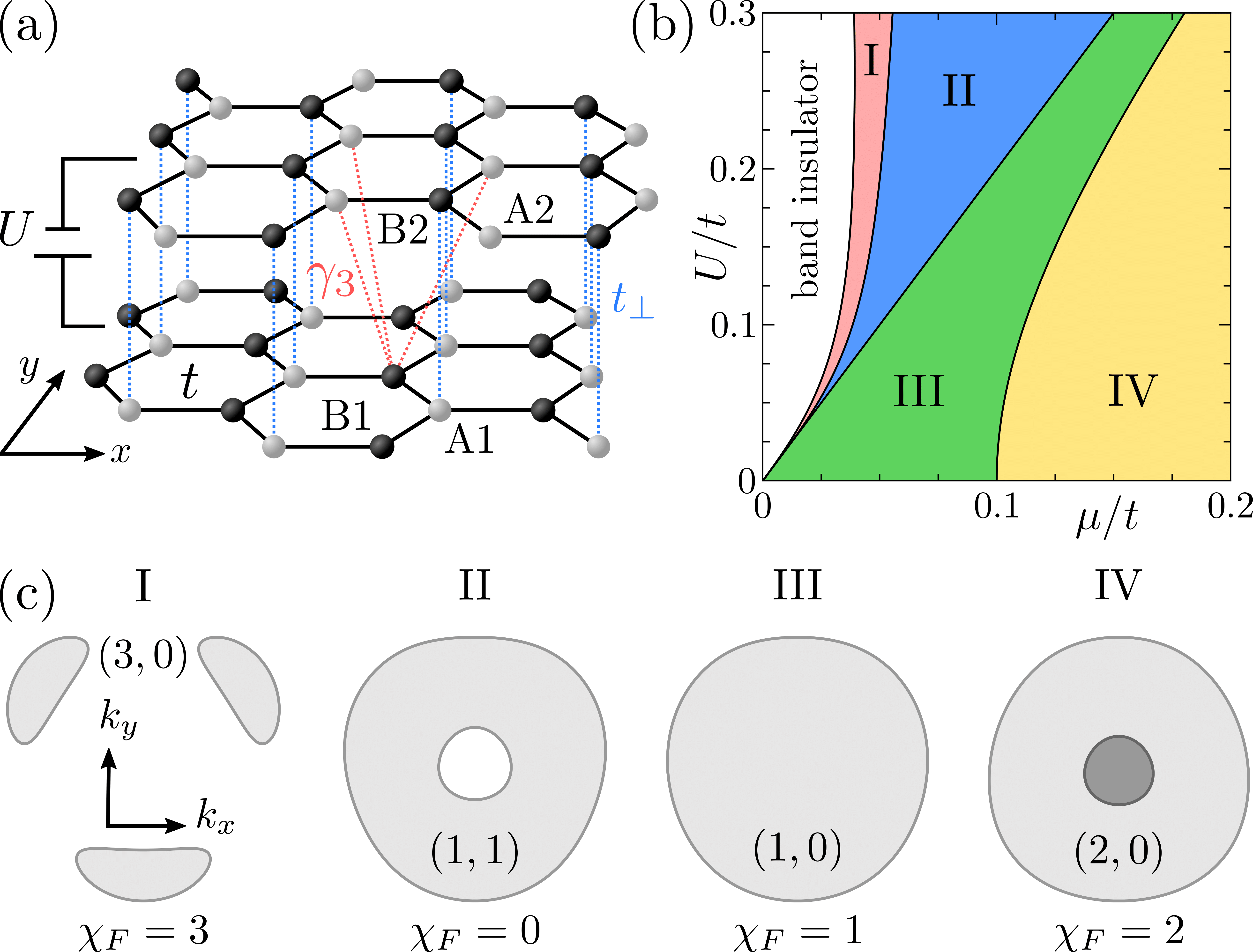}
    \caption{(a) Lattice of Bernal bilayer graphene with interlayer bias $U$ showing the intralayer ($t$) and interlayer ($t_\perp$ and $\gamma_3$) hoppings that we take into account in Eq.\ \eqref{eq:blgH0}. (b) Fermi sea topology $\chi_F$ as a function of the chemical potential $\mu$ and $U$ for $t_\perp=\gamma_3=0.1t$. (c) Fermi sea for the four cases shown in (b). The Fermi surface critical point count $(c_e,c_h)$ relative to the $+\hat{x}$ direction is indicated. In IV the Fermi sea of the second conduction band (darker) overlaps with the lower band. Here $\chi_F$ and $c_{e,h}$ are counted per spin and valley.}
    \label{fig:blg}
\end{figure}

To investigate transport along an Andreev junction in BLG, we use the following lattice model, $\hat{H} = \sum_{l=1,2} \hat{H}_l + \hat H_\perp + \hat{H}_{\Delta}$ with
\begin{widetext}
\begin{align}
    \hat H_l & = \sum_{\bm R,\sigma} \left[ - \mu_l \left( \hat a_{\bm R,l,\sigma}^\dag \hat a_{\bm R,l,\sigma} + \hat b_{\bm R,l,\sigma}^\dag \hat b_{\bm R,l,\sigma} \right) - t \left( \hat a_{\bm R,l,\sigma}^\dag \hat b_{\bm R,l,\sigma} + \hat a_{\bm R+\bm a_1,l,\sigma}^\dag \hat b_{\bm R,l,\sigma} + \hat a_{\bm R+\bm a_2,l,\sigma}^\dag \hat b_{\bm R,l,\sigma} + \text{h.c.} \right) \right], \label{eq:blgH0} \\
    \hat H_\perp & = \sum_{\bm R,\sigma} \left[ t_\perp \left( \hat b_{\bm R,2,\sigma}^\dag \hat a_{\bm R,1,\sigma} + \text{h.c.} \right) + \gamma_3 \left( \hat a_{\bm R+\bm a_1,2,\sigma}^\dag \hat b_{\bm R,1,\sigma} + \hat a_{\bm R+\bm a_2,2,\sigma}^\dag \hat b_{\bm R,1,\sigma} + \hat a_{\bm R+\bm a_1+\bm a_2,2,\sigma}^\dag \hat b_{\bm R,1,\sigma} + \text{h.c.} \right) \right], \\
    \hat H_\Delta & = \sum_{\bm R} \left( \Delta_{\bm R} \hat a_{\bm R,1,\uparrow}^\dag \hat a_{\bm R,1,\downarrow}^\dag + \Delta_{\bm R+\bm \tau} \hat b_{\bm R,1,\uparrow}^\dag \hat b_{\bm R,1,\downarrow}^\dag + \Delta_{\bm R-\bm \tau} \hat a_{\bm R,2,\uparrow}^\dag \hat a_{\bm R,2,\downarrow}^\dag + \Delta_{\bm R} \hat b_{\bm R,2,\uparrow}^\dag \hat b_{\bm R,2,\downarrow}^\dag + \text{h.c.} \right),
\end{align}
\end{widetext}
where $\mu_{1/2} = \mu \pm U/2$ with $U$ the interlayer bias and $\mu$ the chemical potential. Here $\hat a_{\bm R,l,\sigma}^\dag$ and $\hat b_{\bm R,l,\sigma}^\dag$ ($\hat a_{\bm R,l,\sigma}$ and $\hat b_{\bm R,l,\sigma}$) create (destroy) an electron with spin $\sigma = \uparrow,\downarrow$ on layer $l=1,2$ on sublattice A and B, respectively. The in-plane position of sublattices A and B are, respectively, $\bm R$ and $\bm R+\bm \tau$ for layer $1$ and $\bm R - \bm \tau$ and $\bm R$ for layer 2. The most important hoppings \cite{Castro2010} are given by the intralayer nearest-neighbor hopping amplitude $t$ and the interlayer hopping between eclipsing sites on different layers $t_\perp \approx 0.1t$. We also include second-nearest-neighbor interlayer hopping between different sublattices $\gamma_3 \lesssim 0.1t$. The latter gives rise to trigonal warping \cite{Castro2010} and allows for a richer variety of Fermi sea topology. The lattice and hoppings are illustrated in Fig.\ \ref{fig:blg}(a). For simplicity, we take the same value for the proximity-induced pair potential in both layers: $\Delta_{\bm r} = \Delta_0 \left[ \theta(y - W/2) - \theta(-y - W/2) \right]$. The parameters used for the \textsc{Kwant} simulation are given in Table \ref{table:graphene_values}.

\begin{figure}[t]
    \centering
    \includegraphics[width=\linewidth]{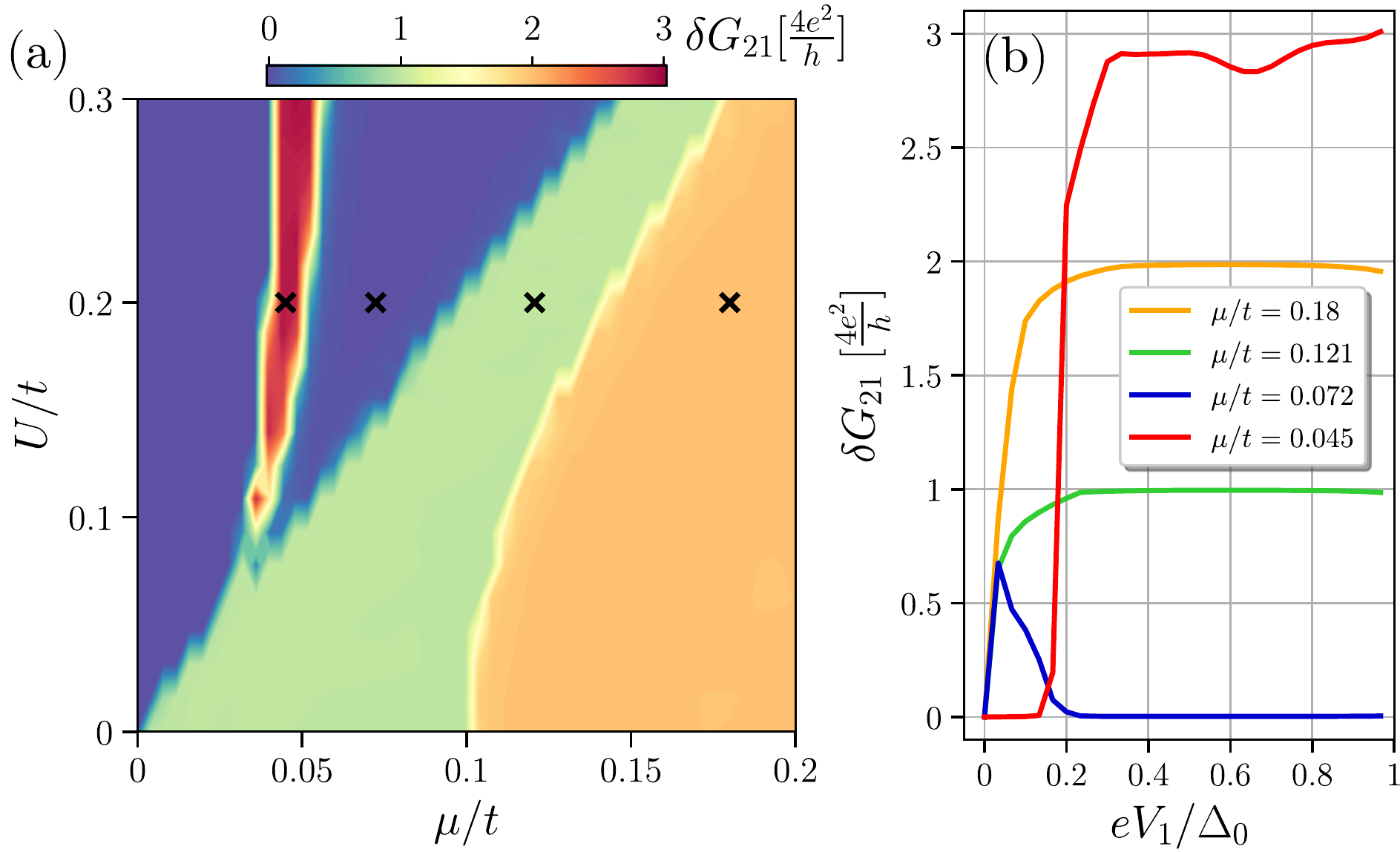}
    \caption{Nonlocal transport for a narrow Andreev junction based on BLG along the armchair direction. The normal region has width $W=4a$ and length $L_x = 600a$, and the scattering region has dimensions $L_x \times L_x$. (a) Rectified nonlocal conductance $\delta G_{21}$ for $eV_1/\Delta_0 = 0.5$ (with $\Delta_0 \approx 7 \; \text{meV}$) as a function of interlayer bias $U$ and chemical potential $\mu$. (b) $\delta G_{21}$ as a function of $eV_1$ for $U=0.2t$ and a set of $\mu$ values, indicated by the crosses in (a).}
    \label{fig:blg-phase}
\end{figure}

The interlayer bias $U$ breaks inversion symmetry and opens a band gap $E_g$ at the $K$ and $K'$ points. In the absence of $\gamma_3$ hopping, the Fermi sea at each valley is given by an annulus for $E_g/2 < |\mu| < U/2$ \cite{Castro2010}. Upon turning on $\gamma_3$, a small energy window appears starting from the band edge where the annular Fermi sea fractures into three electronlike ($\mu>0$) or holelike ($\mu<0$) Fermi pockets. We henceforth focus on $\mu>0$. Owing to a chiral symmetry that relates the conduction and valence band, the results for $\mu<0$ are obtained by reversing the sign of $\delta G_{21}$. Figure \ref{fig:blg}(b) shows the Fermi sea topology as a function of $\mu$ and $U$ where representative examples are illustrated in Fig.\ \ref{fig:blg}(c). For $\mu > U/2$, the Fermi sea (per valley and spin) has the topology of either a disk, where $\chi_F = +1$, or a pair of concentric disks when the second conduction band is also occupied, where $\chi_F = +2$.
To make the transport calculation feasible, instead of scaling the lattice (which gets tricky due to $\gamma_3$), we consider a larger superconducting gap $\Delta_0 = 0.0025t \approx 7 \; \text{meV}$. The superconducting coherence length is then $\xi/a \sim \hbar v_F/\pi \Delta_0 \sim 100$. While this value for $\Delta_0$ is about seven times larger than the proximity-induced gap for Nb or MoRe contacts, our aim is to present a proof of principle for TAR in BLG, which exhibits a rich landscape of Fermi sea topology. Note that our results would be unchanged if we could perform simulations for a larger system and a realistic $\Delta_0$. In fact, assuming a clean system, the quantization of $\delta G_{21}$ would only improve since $\Delta_0$ limits the energy resolution of TAR. We consider here a long and narrow Andreev junction for transport along the armchair direction. Here we take $L_x=600a$ and $W=4a$. Our numerical results for the rectified conductance $\delta G_{21}$ are summarized in Fig.\ \ref{fig:blg-phase}. We see that Fig.\ \ref{fig:blg-phase}(a) matches well to Fig.\ \ref{fig:blg}(b), confirming that $\delta G_{21}$ provides an excellent probe of the Fermi sea topology. The quantized plateaus shown in Fig.\ \ref{fig:blg-phase}(b) further confirm that TAR is in principle observable over a wide range of bias voltages.

Finally, we comment on a potential issue related to the experimental implementation of the Andreev junction in BLG. The application of an interlayer bias across the entire sample (including the regions coupled to the superconductor) by means of an additional top gate can be challenging in practice due to screening by the superconductor. However, it is expected that screening is less important for few-layer van der Waals superconductors such as NbSe$_2$ \cite{Efetov2016}. Given the appropriate device fabrication, we believe that the rich landscape of Fermi sea topology in BLG can indeed be extracted experimentally in a ballistic Andreev junction by measuring $\delta G_{21}$. 

\section{Summary and Outlook} \label{sec:summary}

In this work, we presented both theoretical and numerical studies to consolidate a recent proposal in Ref.\ \cite{Tam2022b} which relates the topology of a two-dimensional Fermi sea to the ballistic transport of Andreev bound states (ABSs) in a superconducting $\pi$ junction. To highlight the significance of transport \emph{along} the junction (i.e., between the two normal leads connected to the two ends of the junction) we termed the proposed setup an \emph{Andreev junction}. Reference \ \cite{Tam2022b} considered the limits of adiabatic and ballistic transport in a narrow Andreev junction of width $W \ll \xi$.
In this case, the junction hosts the minimal number of ABSs and the nonlocal conductance $G_{21}(V_1)$ is quantized to $c_e$ or $c_h$ (in units of $e^2/h$) for subgap bias $eV_1 \gtrless 0$. Here $c_{e}$ ($c_{h}$) counts the number of convex (concave) critical points on the Fermi surface. While $c_{e/h}$ depends on the geometry of the Fermi sea, as well as the orientation of the Andreev junction relative to the Fermi sea, their difference gives the Euler characteristic $\chi_F = c_e-c_h$, which is an intrinsic topological characterization of the Fermi sea. This motivated us to introduce the \emph{rectified} nonlocal conductance $\delta G_{21}(V_1) \equiv G_{21}(V_1)-G_{21}(-V_1)$, which is expected to be quantized to $\chi_F$. We referred to this effect as topological Andreev rectification (TAR). 

To evaluate the feasibility of observing TAR, we considered more general device geometries. We first provided a scattering-matrix analysis which also incorporates \emph{wide} Andreev junctions (i.e., $W \gtrsim \xi$) hosting multiple dispersive ABSs around each Fermi surface critical point. The assumption of adiabaticity is then relaxed to allow for backscatterings among counterpropagating ABSs associated to the same Fermi surface critical point, while scattering between distinct Fermi surface critical points is still assumed to be suppressed. We established that, while $G_{21}$ is no longer quantized for wide junctions, $\delta G_{21}$ remains quantized to $\chi_F$ for small but finite bias, whereas additional quantized plateaus in $\delta G_{21}$ are observed at larger bias. Our theoretical analysis is further supported by numerical calculations using \textsc{Kwant} \cite{codes}. We first simulated nonlocal transport in Andreev junctions with a toy model on a square lattice. Our simulations for a narrow linear junction, as well as for a wide linear junction, confirm the predicted TAR effect. Furthermore, TAR is also demonstrated for an Andreev junction in a \emph{point-contact} geometry. While such Andreev junctions cannot be studied analytically, our numerical results clearly establish TAR as a robust phenomenon insensitive to both real-space and momentum-space microscopic details.

Motivated by recent experimental progress on transport and spectroscopic measurements in Josephson junctions, we also simulated the TAR effect in realistic materials. Specifically, we considered a 2DEG in an InAs quantum well and graphene, for which superconducting junctions have already been fabricated (for InAs, see Refs.\ \cite{Fornieri2019, Banerjee2022_1, Banerjee2022_2, Banerjee2022_3}; for graphene, see Refs.\ \cite{Heersche2007,Du2008,Calado2015,BenShalom2016,Efetov2016,Allen2016,Amet2016,Bretheau2017,Nanda2017,Wang2018,Li2018,Park2022}). Using experimentally relevant parameters, and assuming a clean system, we examined the criteria for TAR to be observed in these systems. For an InAs quantum well proximitized by Al contacts ($\Delta_0 \approx 0.2 \; \text{meV}$) TAR is seen for a junction length $L \gtrsim 3 ~\mu$m when $\mu \sim 10 \; \text{meV}$ (counted from the band bottom). We find that the rectified conductance $\delta G_{21}$ is nearly quantized to $\chi_F = 2$ (in units of $e^2/h$) over a wide range of subgap bias voltages. For monolayer graphene, we considered both Al and Nb (or MoRe) contacts, where the latter induces a larger gap $\Delta_0 \approx 1 \; \text{meV}$. In this case, we considered a device of dimensions $3 \times 3 \; \mu$m$^2$ and observed TAR where $\delta G_{21}$ is now quantized to $\pm 4e^2/h$ for electron or hole doping, respectively. Quantization of $G_{21}$ to $c_e e^2/h$ (for $eV_1 > 0$) and $c_h e^2/h$ (for $eV_1 < 0$), which reflects the Fermi sea geometry, is only obtained with Nb (or MoRe) contacts, owing to a smaller coherence length that puts our setup in the long-junction regime. Finally, for Bernal bilayer graphene, which hosts a rich variety of Fermi sea topologies upon tuning the interlayer bias and chemical potential, we demonstrate once again that $\delta G_{21}$ serves as a good marker of the Fermi sea topology. Our numerical results thus indicate that both InAs quantum wells and graphene are promising candidate platforms for observing topological Andreev rectification.

While our simulations have mostly used experimentally relevant parameter values, we have largely ignored the role of disorder. Our transport simulations are thus in the ballistic regime and perfect quantization in both $G_{21}$ and $\delta G_{21}$ can be achieved for $L \gg \xi$, in accordance with theoretical predictions. However, the current experimental status suggests that the electron mean free path $\ell_e$ is of the order of $\xi$. We therefore believe that a major challenge for observing quantization in both $G_{21}$ and $\delta G_{21}$ lies in increasing the mobility in these platforms such that $\ell_e$ becomes at least \emph{one order longer} than $\xi$. Importantly, however, achieving quantization in $\delta G_{21}$ (i.e., TAR) is much easier than quantization in $G_{21}$. As indicated by our simulations, TAR is observed for shorter junctions for which $G_{21}$ is not quantized. Furthermore, a \emph{linear} junction geometry is \emph{not required}, as TAR has also been consistently demonstrated for Andreev point contacts. 

Finally, even if \emph{quantized} topological Andreev rectification cannot be observed in existing devices due to disorder, as long as the variation in chemical potential across the sample is not so drastic to create multiple regions with distinct Fermi sea topology (e.g., for graphene we would require the absence of electron-hole puddles), the signatures of Andreev rectification would still remain for $\chi_F \neq 0$. In particular, $\delta G_{21}$ is expected to change drastically as the quantized value of $\chi_F$ jumps under a Lifshitz transition of the underlying metal in the Andreev junction. In short, we believe that the topological Andreev rectification effect is experimentally testable in the near future. 

\begin{acknowledgments}
We thank M.\ Claassen for computational resources. This research was funded in whole, or in part, by
the Luxembourg National Research Fund (FNR) (project
No. 16515716). Work by P.M.T.\ and C.L.K.\ was supported by a Simons Investigator
Grant to C.L.K.\ from the Simons Foundation. 
\end{acknowledgments}

\appendix

\section{Landauer-B\"uttiker formalism} \label{app:landauer}

We consider a four-terminal setup with two normal leads (leads N1 and N2) and two superconducting leads (leads S1 and S2) as illustrated in Fig.\ \ref{fig:system} of the main text. Here the bias at the normal leads is given by $\mu_1=eV_1$ and $\mu_2=eV_2$ while the superconducting leads are grounded with $\mu_S = 0$. When transport is coherent across the system, the current that flows \emph{into} the normal leads is given by \cite{Takane1992, Datta1996scattering}
\begin{widetext}
\begin{align}
    I_{1} & = -\frac{e}{h} \int_{-\infty}^\infty d\varepsilon \left\{ \left( M_1^e - T_{11}^{ee} + T_{11}^{he} \right) \left[ f_0(\varepsilon-\mu_{1}) - f_0(\varepsilon) \right] + \left( T_{12}^{he} - T_{12}^{ee} \right) \left[ f_0(\varepsilon-\mu_{2}) - f_0(\varepsilon) \right] \right\}, \\
    I_{2} & = -\frac{e}{h} \int_{-\infty}^\infty d\varepsilon \left\{ \left( M_2^e - T_{22}^{ee} + T_{22}^{he} \right) \left[ f_0(\varepsilon-\mu_{2}) - f_0(\varepsilon) \right] + \left( T_{21}^{he} - T_{21}^{ee} \right) \left[ f_0(\varepsilon-\mu_{1}) - f_0(\varepsilon) \right] \right\},
\end{align}
\end{widetext}
with $e = -|e|$  the electron charge and $f_0(\varepsilon) = 1 / \left[ \exp \left( \varepsilon / k_B T \right) + 1 \right]$ the Fermi-Dirac distribution function, where the excitation energy $\varepsilon$ is defined relative to the equilibrium chemical potential $\mu$. Here we suppressed the subscript $N$ in the total (including spin) transmission functions between the normal leads
\begin{equation}
    T_{ij}^{\alpha\beta}(\varepsilon) = \mathrm{Tr}\left[ ( s_{ij}^{\alpha\beta} )^\dag s_{ij}^{\alpha\beta} \right],
\end{equation}
with $i, j = 1, 2$ and $\alpha, \beta = e, h$. The matrices $s_{ij}^{\alpha\beta}$ contain the amplitudes for scattering from $\beta$-type incident modes in lead N$j$ to $\alpha$-type outgoing modes in lead N$i$. In the subgap regime $|\varepsilon|<\Delta_0$, the total number of electron modes in the normal leads is given by $M_i^e = \sum_j \left( T_{ij}^{ee} + T_{ij}^{eh} \right)$. Note that electron-hole symmetry implies $T_{ij}^{ee}(\varepsilon) = T_{ij}^{hh}(-\varepsilon)$ and $T_{ij}^{eh}(\varepsilon) = T_{ij}^{he}(-\varepsilon)$, while reciprocity implies $T_{ij}^{\alpha\beta}(\varepsilon) = \overline T_{ji}^{\beta\alpha}(\varepsilon)$ where $\overline T$ is calculated for the time-reversed system. The local and nonlocal differential conductance are then, respectively,
\begin{align}
    G_{11} & = \frac{dI_1}{dV_1} = -\frac{e^2}{h} \int_{-\infty}^\infty d\varepsilon \left( M_1^e - T_{11}^{ee} + T_{11}^{he} \right) \left( - \frac{df_1}{d\varepsilon} \right), \\
    G_{21} & = \frac{dI_2}{dV_1} = \frac{e^2}{h} \int_{-\infty}^\infty d\varepsilon \left( T_{21}^{ee} - T_{21}^{he} \right) \left( - \frac{df_1}{d\varepsilon} \right),
\end{align}
where $f_1(\varepsilon) = f_0(\varepsilon - \mu_{1})$ with $\mu_{i} = eV_i$ ($i=1,2$). In the zero-temperature limit, we obtain
\begin{align}
    \lim_{T \rightarrow 0} G_{11} & = -\frac{e^2}{h} \left( M_1^e - T_{11}^{ee} + T_{11}^{he} \right)_{\varepsilon = eV_1}, \\
    \lim_{T \rightarrow 0} G_{21} & = \frac{e^2}{h} \left( T_{21}^{ee} - T_{21}^{he} \right)_{\varepsilon = eV_1}.
\end{align}
The rectificed nonlocal conductance is
\begin{equation}
\begin{split}
    \delta G_{21}(V_1) & = G_{21}(V_1) - G_{21}(-V_1) \\
    {\scriptstyle (T \rightarrow 0)} & = \frac{e^2}{h} \left( T_{21}^{ee} - T_{21}^{hh} - T_{21}^{he} + T_{21}^{eh} \right)_{\varepsilon=eV_1} \\
    & = \frac{e^2}{h} \left( T_{21}^{ee} - T_{21}^{hh} - T_{21}^{he} + \overline T_{12}^{he} \right)_{\varepsilon=eV_1},
\end{split}
\end{equation}
where electron-hole symmetry and reciprocity have been used. For a transport setup that is both time-reversal symmetric and mirror symmetric along the transport direction (i.e., under $\mathcal M_x: \; x \mapsto -x$ for the setup shown in Fig.\ \ref{fig:system} of the main text), we have $\overline T_{12}^{he} = T_{12}^{he} = T_{21}^{he}$, and hence contributions from crossed Andreev reflections are canceled. The quantization in $\delta G_{21}$, as established in Sec.\ \ref{sec:smatrix}, still relies on having a long junction with $L \gg \xi$ so that $T^{ee}_{21}$ and $T^{hh}_{21}$ only contain contributions from the dispersive ABSs (and not from tunneling).

Let us also compute the rectified \emph{local} conductance $\delta G_{11}$ here, which is not studied explicitly in the main text. Using $M^e_1 = T^{ee}_{11}+T^{he}_{11}+T^{ee}_{21}+T^{he}_{21}$ and electron-hole symmetry, we have
\begin{equation} \label{eq:dg11}
\begin{split}
    \delta G_{11}(V_1)  &= G_{11}(V_1) - G_{11}(-V_1) \\
    {\scriptstyle (T \rightarrow 0)} & = -\frac{e^2}{h} \big( 2T^{he}_{11}-2T^{eh}_{11}+T^{ee}_{21}-T^{hh}_{21}\\
    &\quad\quad\quad\quad\quad\quad+T^{he}_{21}-T^{eh}_{21} \big)_{\varepsilon=eV_1}.
\end{split}
\end{equation}
As before, reciprocity and time-reversal symmetry $\mathcal T$ give $T^{he}_{11} = \overline T^{eh}_{11} = T^{eh}_{11}$, such that the first two terms in Eq.\ \eqref{eq:dg11} cancel each other. Hence, when $\mathcal T$ is preserved, we have
\begin{align}
    \delta G_{21}(V_1) + \delta G_{11}(V_1) & = \frac{2e^2}{h} \left( T_{21}^{eh} - T_{21}^{he} \right), \label{eq:sumdg} \\
    \delta G_{21}(V_1) - \delta G_{11}(V_1) & = \frac{2e^2}{h} \left( T_{21}^{ee} - T_{21}^{hh} \right). \label{eq:diffdg}
\end{align}
When, in addition to $\mathcal T$ symmetry, $\mathcal M_x$ is preserved, we have $T^{eh}_{21} = T ^{he}_{21}$ and hence $\delta G_{11} = -\delta G_{21}$. If $\mathcal M_x$ is not present, but the junction is long enough such that crossed Andreev reflection is suppressed, we also have $\delta G_{11} \approx -\delta G_{21}$. Moreover, when $\mathcal T$ symmetry is broken but $\mathcal M_y \mathcal T$ is conserved, using reciprocity, the relations given in Eq.\ \eqref{eq:sumdg} and Eq.\ \eqref{eq:diffdg} still hold.

In this work we have focused on the nonlocal conductance $G_{21}$ instead of the local conductance $G_{11}$, as not only is $\delta G_{21}$ quantized to reflect the Fermi sea topology, in the case of a narrow Andreev junction $G_{21}$ is \emph{itself} also quantized to reflect the Fermi sea geometry.

\end{document}